\documentclass[floatfix,prl,twocolumn]{revtex4-1} 
\usepackage{epsfig} 
\usepackage{amssymb,amsmath}

\usepackage{graphicx} 
\usepackage{dcolumn}
\usepackage[normalem]{ulem}
\usepackage{lipsum}
\usepackage{verbatim}
\usepackage{color}

\begin{document}

\title{Strongly correlated two-dimensional plasma explored from entropy measurements}

\author{A.~Yu.~Kuntsevich} \affiliation{Lebedev Physical Institute of the RAS, 119991 Moscow, Russia}.  \affiliation{Moscow Institute of Physics and
Technology, 141700 Moscow, Russia}

\author{Y.~V.~Tupikov} \affiliation{Department of Physics, Pennsylvania State University, University Park, PA 16802, USA}

\author{V.~M.~Pudalov} \affiliation{Lebedev Physical Institute of the RAS, 119991 Moscow, Russia} \affiliation{Moscow Institute of Physics and Technology,
141700 Moscow, Russia}

\author{I.~S.~Burmistrov} \affiliation{L.D. Landau Institute for Theoretical Physics, Kosygina
  street 2, 119334 Moscow, Russia}
\affiliation{Moscow Institute of Physics and Technology, 141700 Moscow, Russia}

\begin{abstract}

Charged plasma and Fermi liquid are two distinct states of electronic matter intrinsic to dilute two-dimensional electron systems at elevated and low temperatures, respectively. Probing their thermodynamics represents challenge because of lacking an adequate technique.
Here we report thermodynamic method to measure the
entropy per electron in gated structures.  Our technique appears to be
three orders of magnitude superior in sensitivity to the ac calorimetry,
allowing entropy measurements with only 10$^8$ electrons. This enables us to
investigate the correlated plasma regime, previously inaccessible
experimentally in two-dimensional electron systems in semiconductors. In
experiments with clean two-dimensional electron system in Si-based
structures we traced entropy evolution from the plasma to Fermi-liquid
regime by varying electron density. We reveal that the correlated plasma
regime can be mapped onto the ordinary non-degenerate Fermi gas with an
interaction-enhanced temperature dependent effective mass. Our method
opens up new horizons in studies of low-dimensional electron systems.
\end{abstract}

\date{\today} \maketitle
Interaction between conduction electrons is well known to become more and more important as dimensionality  is reduced and as carrier density
decreases. Consequently, the dilute two dimensional (2D) electron systems exhibit a variety of  spectacular interaction effects  such as the fractional quantum Hall effect \cite{nobel_lecture}, metal-insulator transition (MIT) in
2D \cite{kravchenko,finkelstein_Science},  strong enhancement in the effective mass and spin susceptibility \cite{pudalovFL, clarke-natphys_2008}, etc. 
The vast majority of these phenomena are explored using charge transport and Coulomb drag\cite{drag} measurements. The apparent simplicity of these techniques, however, is spoiled by an indirect character of the results whose interpretation requires specific theoretical models. In contrast, thermodynamic measurements  carry  more direct information on the electron-electron correlations. However,  these measurements are
harder because of the smallness of thermodynamic quantities for a dilute single-layer 2D system confining, typically,  $10^8 - 10^{10}$ electrons.

By now, only electron compressibility \cite{eisenstein_PRL_1992, allison, dultz},  spin magnetization \cite{tenehdmudb}, and chemical potential \cite{pudalov_1986} have been  measured in low density 2D carrier systems, whereas the major thermodynamic parameter - entropy - has never been explored.

\begin{figure*}
\vskip.05in
\begin{center}
\centerline{\psfig{figure=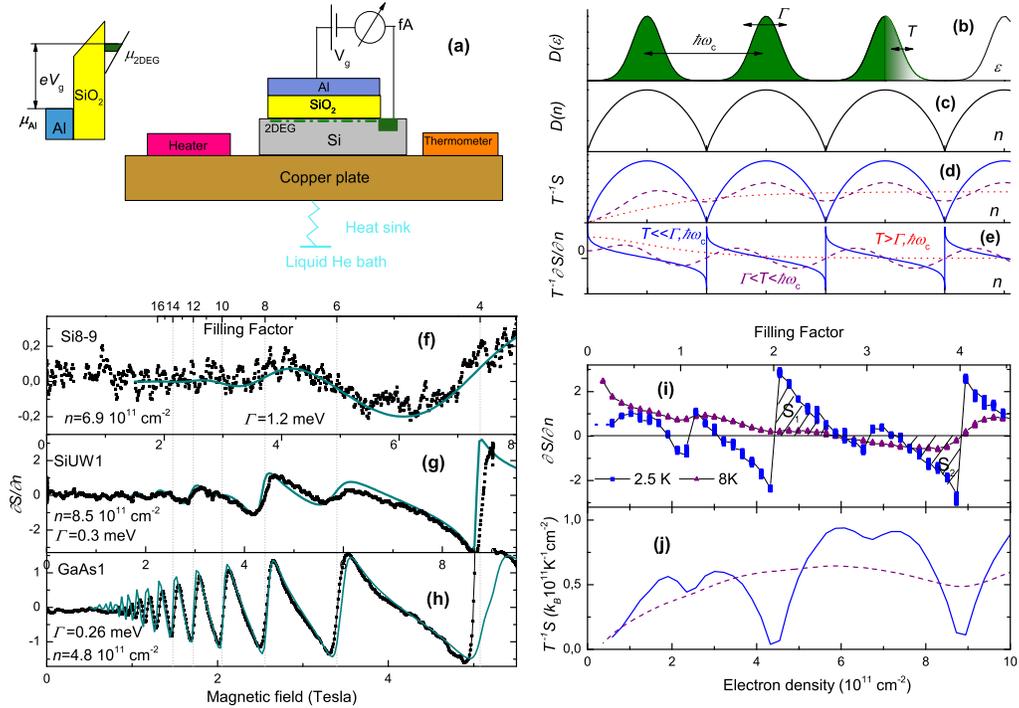, width=400pt}}
\caption{
{\bf Scheme of experiment and and entropy in perpendicular magnetic field.}
 {(a) Experimental setup and Si-MOSFET-structure. (b--c) The density of states for a 2D electron gas in quantizing magnetic field versus energy and electron density,
 respectively. Green color denotes filled states.
 (d--e) Ratios $S/T$ and  $T^{-1}\partial S/\partial n$ versus electron density for different relationships between  $\Gamma , T$ and $\hbar\omega_c$ (shown in the figure).
 (f--h) The entropy per electron at $T=3$\,K versus magnetic field measured with three samples ((f) Si8-9, (g) SiUW1, and (h) GaAs1). The upper axis (filling factor $\nu=e B/(h n)$ ) is common for all three panels. The densities are shown on the panels. Cyan curves represent the fit using the non-interacting electron gas model with the band effective masses and level broadenings indicated on the panels.}
 (i) $\partial S/\partial n$ versus density measured at 9\,T
 with SiUW2. The temperatures are shown on the panel. Hatched regions S$_1$ and S$_2$ have approximately the same area (see text). The data spread for each density at 2.5 K indicates typical noise level. (j) Normalized entropy versus density for the same sample.
}
\label{Fig0}
\end{center}
\end{figure*}

Here we show how
the entropy per electron, $\partial S/\partial n$,  can be measured
for
2D electron systems  ($n$ is the  electron density).
According to the Maxwell relation $(\partial S/\partial n)_T=-(\partial \mu/\partial T)_n$, the
sought for entropy is found from the temperature derivative of the chemical potential. In order to measure $\partial \mu/\partial T$  of the 2D electron layer, we  use a gated structure
(shown schematically in Fig.~\ref{Fig0}a) where the 2D electrons and the gate act as two plates of a capacitor. When temperature is
modulated, the chemical potential of the 2D layer varies causing the capacitor recharging.

{\bf Results}

{\bf The experimental set-up}
is similar to the one used in Ref.~\cite{tenehdmudb} for spin magnetization measurements.
The sample (either Si-MOSFET, or GaAs-FET) and the copper sample holder are kept in a thermal contact with a wire heater which modulates their temperature:
$T(t)=T_0+\Delta T \cos(\omega
t)$. In our experiments, temperature $T_0$ typically varies from  2.4 to 26\,K, and is modulated at the frequency $\omega/(2\pi) \approx 0.5$\,Hz with an
amplitude $\Delta T \sim 0.1-0.2$\, K. See Supplementary Note 1 for details concerning temperature modulation.

Modulation of the sample temperature changes the chemical potential and, hence, causes recharging of the gated structure. Therefore, $\partial\mu/\partial T$ is directly determined in the experiment from the measured recharging current:
 \begin {equation}
 i(t) = \frac{\partial \mu}{\partial T}\, \Delta T \, \omega\, C \sin(\omega t) .
 \label{experimental}
 \end{equation}
Here $C$ stands for the capacitance between the gate electrode and 2D electron layer. Both $T_0$ and $\Delta T$ values were measured  using a miniature thermometer
attached to the copper sample chamber, and $C$ was determined in the same experiment. Using the PPMS-9 cryomagnetic system we  applied magnetic field up to 9 T either parallel or perpendicular to the sample plane.We used the following four samples: a Schottky-gated GaAl/AlGaAs heterostructure (sample GaAs1) with  peak mobility about 20\, m$^2$V$^{-1}$s$^{-1}$, and three Si-MOS structures (SiUW1, SiUW2 and Si8-9) with the peak mobilities  about 3\, m$^2$V$^{-1}$s$^{-1}$, 3\, m$^2$V$^{-1}$s$^{-1}$ and 0.5\, m$^2$V$^{-1}$s$^{-1}$, respectively. For more details concerning the introduced measurement technique see Supplementary Note 2.

{\bf Measurements in quantizing magnetic fields.}
In order to test the technique, we start
 from measurements of $\partial S/\partial n$   in quantizing perpendicular magnetic fields. In  strong fields the energy spectrum of the system consists of a ladder of the broadened Landau levels whose density of states $D(\varepsilon)$ is shown schematically in  Fig.~\ref{Fig0}b. For simplicity, we focus on the cyclotron splitting $\hbar \omega_c$ and neglect Zeeman splitting, and interaction effects.
In the low-temperature limit ($T\ll \hbar\omega_c, \Gamma, E_{\rm F}$)  the entropy of the noninteracting 2D electron gas is given by
\begin{equation}
S=\pi^2 T D/3,
\label{entropy}
\end{equation} where  $D=D(\varepsilon=E_{\rm F})$ is  the
 density of states at the Fermi energy.
We use throughout the paper $k_{\rm B}=1$. In order to find $\partial S/\partial n$
we express $D$ versus $n$ (see Fig.~\ref{Fig0}c) using the definition $n=\int D(\varepsilon) f_{\rm F}(\varepsilon) d\varepsilon$ with $f_{\rm F}(\varepsilon)=1/[1+\exp{((\varepsilon-\mu)/T})]$ being the Fermi-Dirac distribution function.
The resulting anticipated behavior of the entropy and  $\partial S/\partial n$ as a function of $n$ is plotted in Figs.~\ref{Fig0}d and \ref{Fig0}e, respectively.

We emphasize that the calculated entropy vanishes together with $D$ in the gaps between Landau levels. As temperature increases and becomes larger than the level broadening $\Gamma$, the  entropy dips get smeared (see Figs. \ref{Fig0}d,e) being averaged over energy interval $\sim T$. Finally, in the
high temperature limit $T\gg\hbar\omega_c$, the level quantization is washed out and the entropy $S(n)$ should approach the monotonic dependence anticipated for zero magnetic field (see below).

 In  Figs.~\ref{Fig0}f,g,h the calculated entropy is compared with the measured magnetic field dependence of $\partial S/\partial n$ for three different samples. The cyan curves show a fit ( for details of the fit see Supplementary Note 3) of the data with the theoretical model (given in Supplementary Note 4) for noninteracting electrons using  a bare band mass and a realistic level broadening. The data for the most disordered sample Si8-9 demonstrates only cyclotron gaps resolved at the filling factors $\nu=4$ and $\nu=8$ ( for (100) Si-MOS structures, the 2D electron layer has additional twofold valley degeneracy). In order to fit the data at $\nu=6$ (spin-gap) for the clean sample SiUW1 we used an enhanced g-factor $g_{\rm eff}=4$. As known, enhancement of the spin gap due to the inter-level exchange interaction is a precursor of the quantum Hall ferromagnetism \cite{Girvin}.

 The good agreement of the data for GaAs1 sample  with the noninteracting theory (Fig.~\ref{Fig0}h),  allows us to compare our method with the state-of-the-art ac calorimetry performed with similar GaAs-samples \cite{wang}. It appears that our technique has three orders of magnitude higher resolution than the ac calorimetry (see Supplementary Note 5). Importantly, it is also free from the lattice specific heat contribution.

We now consider the interacting 2D electron system in Si-MOSFET samples. Since the density of states in the middle of the gap (at $\nu=2$ or $\nu=4$) is negligibly small, the entropy at both points should be close to zero. This implies that the area $S_1$ (positive signal) and area $S_2$ (negative signal) should  completely  compensate each other,
as shown in Fig.~\ref{Fig0}i. This integral property is not affected by correlation effects and is fulfilled in the experimental data with a good accuracy.
We can now evaluate the entropy of the system:
\begin{equation}
 S(n)=\int_{n_0}^n \frac{\partial S}{\partial n} dn + S(n_0).
\label{entrIntegral}
\end{equation}

In the ideal case, for integration one would have to choose $n_0=0$ where $S(0)=0$ by definition. However, the 2D system can never be fully depleted:
below a threshold density $n_{\rm th}$ it becomes nonconducting and fails to recharge. For our temperature range $T>2.5K$ and in magnetic fields $B<1$ T we could deplete the samples SiUW1 and SiUW2 down to densities $n_{\rm th}\sim 
3 \cdot  10^{10}$\,cm$^{-2}$ which correspond to $\sim
1$\,GOhm  resistance of the 2D electron system. Therefore in a strong magnetic field it is more practical to choose  $n_0$ as the initial point for the integration such that it corresponds to an even
filling factor ($\nu=2$ or $4$ in our case) where the gap is large. This choice ensures  the  $S(n_0)$ value to be close to zero.

The integrated entropy is shown in Fig.~\ref{Fig0}j. As expected from Fig.~\ref{Fig0}i, at even filling factors ($\nu=0,2,4$) the entropy is almost zero. This fact supports our method of entropy reconstruction.

\begin{figure*}
\begin{center}
\centerline{\psfig{figure=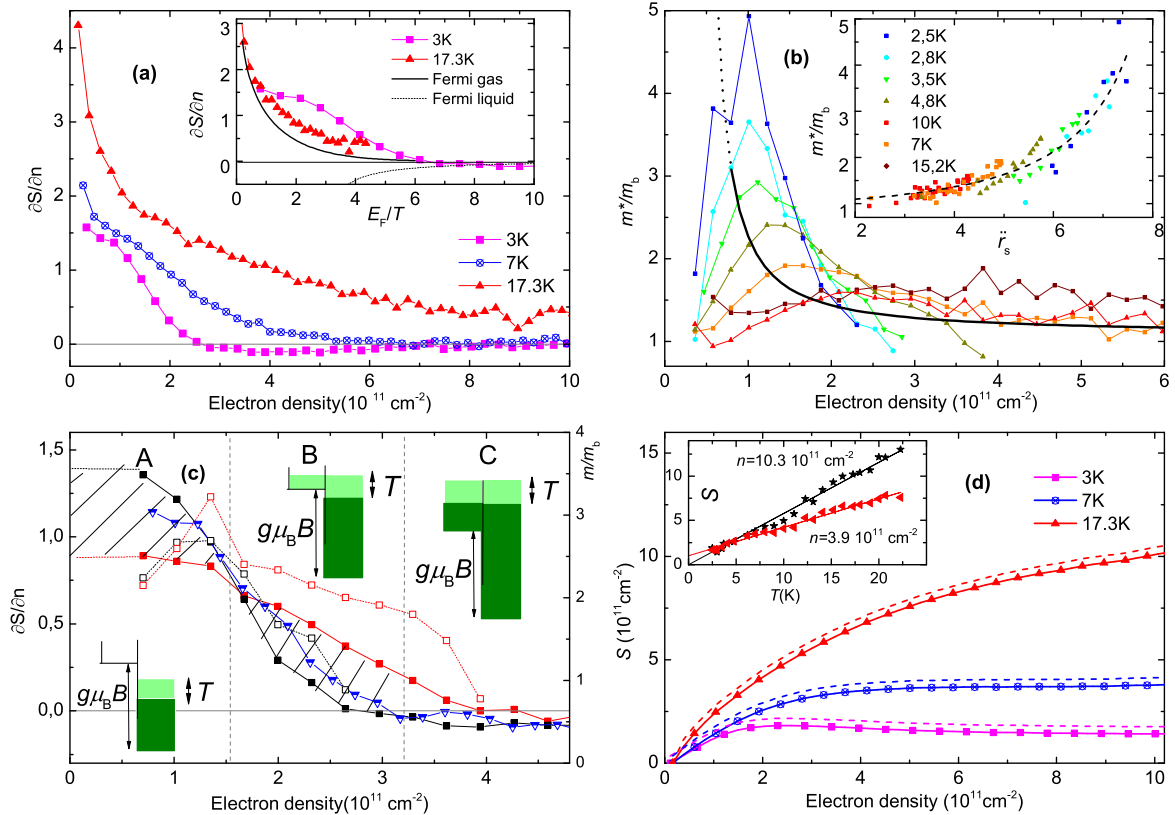, width=450pt}}
\caption{ {\bf Entropy per electron in zero perpendicular field.}(a) The entropy per electron $\partial S/\partial n$ versus density (symbols)
for various temperatures, sample SiUW2. Inset: the same data versus dimensionless density ($E_{\rm F}/T$), the solid curve is the expectation for the Fermi gas with the Si band parameters, the dashed curve is the expectation for the FL with negative $\partial m^*/\partial n$ (see text);
(b) the effective mass $m^*$ versus density. The black curve corresponds to the approximation of $m^*$ from the Shubnikov-de-Haas measurements  Ref.~\cite{pudalovFL}. Symbols are the $m^*(n,T)$ data determined using Eq.~(\protect{\ref{idealFG}}) from the measured $\partial S/\partial n$ values. Different symbols correspond to different temperatures (shown in the inset). Scaling of the effective mass versus effective interaction parameter $\tilde{r}_s$ is shown in the inset(see text).
(c) The signal $\partial S/\partial n (n)$ at 3.2K for SiUW2 is shown with filled symbols: at zero field (black boxes), $B_\parallel=5.5$\,T (blue triangles) and 9\,T (red boxes). Empty symbols (right axis) are the corresponding effective masses at $B=0$ (black) and $B_\parallel=9$\,T (red). The bars illustrate schematically the band diagram for two spin subbands in the regions $\mathbf{A}$, $\mathbf{B}$, and $\mathbf{C}$.
Vertical dashed lines depict schematic borders between the regions  $\mathbf{A}$, $\mathbf{B}$, and $\mathbf{C}$.
(d) The entropy of the 2D electron system measured in SiUW2 for three temperatures (symbols). Inset: temperature dependence of the entropy for $n=10.5\cdot  10^{11}$ cm$^{-2}$ ($E_{F}=75$K) and  $3.9\cdot 10^{11}$ cm$^{-2}$ ($E_{F}=30$K). Dashed curves denote the upper estimate for the entropy.}
\label{Fig1}
\end{center}
\end{figure*}

{\bf Measurements in zero perpendicular field.}
We now turn to more delicate thermodynamic effects for the 2D electron  system in zero perpendicular field. For noninteracting  2D electron gas with a simple parabolic spectrum ($\varepsilon=p^2/2m_{\rm b}$) the density of states is constant $D=g_{s,v} m_{\rm b}/2\pi\hbar^2$.  Here  $g_{\rm s,v}$ denotes the total (spin and valley) spectrum degeneracy, and $m_{\rm b}$ is the band mass  (for (100)Si-MOS $g_{\rm s,v}=4$, and $m_{\rm b}\approx 0.2m_{\rm e}$ \cite{ando_review}).
In the degenerate limit $E_{F}=D n\gg T$, the entropy $S=\pi T g_{\rm s,v} m_{\rm b}/6\hbar^2$ is independent of density, $\partial S/\partial n=0$,  with the exponential accuracy. In the opposite limit $E_{\rm F}\ll T$, the temperature is the dominant
parameter and the entropy per electron agrees with that for the ideal Boltzman gas $\partial S/\partial n \propto \ln(n/T)$. For arbitrary temperature, the
well-known result  for a noninteracting 2D electron gas  is given as
\begin{equation}
\left (\frac{\partial S}{\partial n}\right )_T=\frac{E_{\rm F}/T}{e^{E_{\rm F}/T}-1}-\ln(1-e^{-E_{\rm F}/T}).
\label{idealFG}
\end{equation}

With high-mobility Si-MOSFET samples, in contrast to the ubiquitous charge  transport,
our thermodynamic technique provides an access to the very dilute regime $n \sim
(3\times 10^{10} - 10^{11}$)\,cm$^{-2}$. For such low densities and  interaction enhanced quasiparticles effective mass $m^*\sim (0.4 - 0.6)m_{\rm e}$   one can readily enter the non-degenerate regime $T> E_{\rm F}$. Moreover,  in this regime  the typical electron-electron interaction energy $U = e^2 \sqrt{n}/\kappa$ is about several
meV, i.e. an order of magnitude higher than both, temperature and the Fermi energy ($\kappa \approx 7.7$ is the effective dielectric constant for the Si/SiO$_2$ interface)
\cite{ando_review}. Such strong interaction modifies the spectrum and brings us to an unexplored \cite{novikov} regime of a correlated 2D charged plasma.

In the low temperature Fermi-liquid (FL) regime, the zero-field entropy is given by Eq.~\eqref{entropy} in which $D$ is
replaced by the density of quasiparticle states at the
Fermi level, $D^*= g_{\rm s,v}m^*/(2\pi\hbar^2)$ \cite{LL9}. In contrast to the
noninteracting case, $D^*$ depends on the electron density
$n$ via the effective mass $m^*(n)$ which includes renormalization
due to electron-electron interaction. Then one finds from Eq.~\eqref{entropy}\ that $\partial S/\partial n=[\pi T g_{\rm s,v}/(6\hbar^2)]\cdot\partial m^*/\partial n$; this relates the measured quantity to the $\partial m^*/\partial n$.
In this regime, the entropy per electron is expected to be negative (see insert to Fig. \ref{Fig1}a) since electron-electron interaction increases the effective mass, $\partial m^*/\partial n < 0$ \cite{pudalovFL}.

 As we found, the mass $m^*$ extracted from our data in the FL regime agrees reasonably with the effective mass measured from damping of the amplitude of quantum oscillations \cite{pudalovFL} (for more detail, see Supplementary note 6).

 As  density decreases, the electron system inevitably enters the correlated plasma regime $E_{\rm F}\lesssim T< U$ for which the FL theory is inapplicable.
There were experimental indications (see \cite{pudalov-spinless, shashkinMindep}, and Figs 3a,b of Ref.~\cite{pudalovFL}) that electron-electron interactions preserve parabolic spectrum $\varepsilon=p^2/2m^*$ in the 2D Fermi-liquid in the wide energy range, rather than in the vicinity of the Fermi-level. We expect therefore that even in the plasma regime the entropy should carry signatures of the interaction-enhanced effective mass.

Examples of the $\partial S/\partial n$  density dependence  for several temperatures are shown in  Fig.~\ref{Fig1}a. Qualitatively, the data is consistent with the above expectations: (i) the larger the temperature, the higher is the entropy (area under the curve); (ii) as the Fermi energy (electron density) increases, $\partial S/\partial n$ tends  to zero, and (iii) for the lowest temperatures and highest densities, $\partial S/\partial n$ becomes negative, in accord with that anticipated for the Fermi liquid.
 The inset of Fig.~\ref{Fig1}a  shows that the experimental $\partial S/\partial n$ data are systematically higher than the  theoretical dependence for noninteracting Fermi gas, calculated from Eq. \eqref{idealFG} (with the bare band mass 0.19 $m_{\rm e}$). Since the entropy  is proportional to the density of states averaged over the energy interval $T$, we attribute such excessive entropy regions to the modified effective mass. Lacking a microscopic theory for non-degenerate strongly interacting electron system, we fitted our results using Eq.~\eqref{idealFG} with a density dependent effective mass $m^*(n)$ as a fitting parameter.

 Thus extracted effective mass for different temperatures is shown in Fig.~\ref{Fig1}b.
  In the high temperature limit $T\gtrsim U\gg E_{\rm F}$ the kinetic energy of electrons is given by temperature; hence, the 2D electron gas turns out to be weakly interacting, and
 $\partial S/\partial n$ is expected to be described by Eq.~\eqref{idealFG} with the density-independent effective mass close to the band mass value $m_{\rm b}$.

In general, for a given temperature the effective mass exhibits a reentrant behavior: as density decreases  $m^*$  first grows, then passes through a maximum, and falls down approaching a value of the order of $m_{\rm b}$. The lower the temperature, the higher maximum value the effective mass reaches. The enhanced effective mass is in a qualitative agreement with the low-temperature Shubnikov-de Haas  measurements of Ref. \cite{pudalovFL} (shown with a thick curve in Fig.~\ref{Fig1}b).

\begin{figure}
\begin{center}
\centerline{\psfig{figure=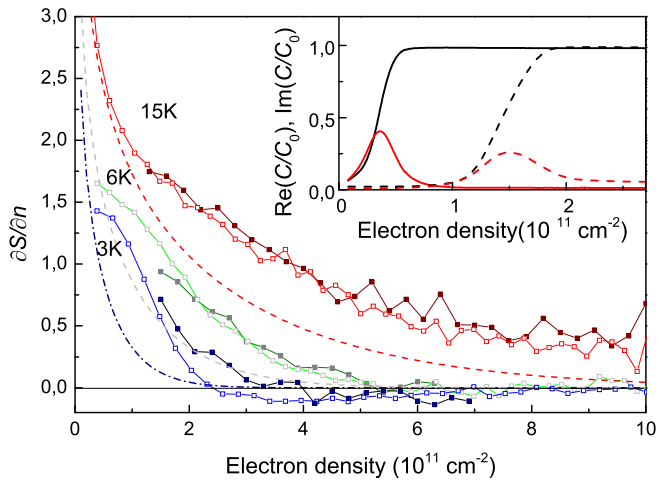, width=270pt}}
\caption{{\bf Comparison of clean and disordered Si samples.} Entropy per electron versus electron density in Si8-9 (filled symbols) and SiUW2 (empty symbols) for 3 different temperatures (15K - red colors, 6K - gray colors, 3K - blue colors). Theory expectation for ideal Fermi gas are shown by red dashed line(15 K), grey dotted line (6K) and blue dash-dotted line(3K). The inset shows real(black) and imaginary(red) components of capacitance taken at the $\partial S/\partial n$ measurement frequency for the clean sample Si-UW2 (solid lines) and disordered sample Si8-9 (dashed lines) respectively at 3K. }
\label{compar}
\end{center}
\end{figure}

The low density region, where the effective mass falls as density decreases, corresponds to a non-degenerate strongly correlated electron plasma regime ($E_{\rm F}\lesssim T<U$). We are not aware of any theory describing this domain. In order to treat the data in this regime we suggest the following phenomenological approach. For the degenerate clean  2D Fermi liquid, renormalization of its physical parameters, and, particularly, the effective mass,  is governed by a single dimensionless variable equal to the ratio of the potential interaction energy $U$ to the kinetic Fermi energy, $r_{\rm s}=1/(\overline{a_{\rm B}^*}\sqrt{\pi n}) \propto U/E_{\rm F}$ \cite{ando_review}. Here $\overline{a_{\rm B}^*}=\overline{\kappa} \hbar^2/(m_{\rm b} e^2)$ stands for the effective Bohr radius with average dielectric constant.

As explained above, when temperature increases, the interactions for a given density weaken and can not be characterized anymore by  $r_{\rm s}\propto U/E_{\rm F}$.
 Correspondingly, to describe  our $m^*(n,T)$ data set over the wide range of densities and temperatures,  we suggest a phenomenological effective interaction parameter $\tilde{r}_{\rm s}=(\pi a_{\rm B}^2 n +\alpha T^{\gamma+\beta}/E_{\rm F}^\gamma U^\beta)^{-1/2}$ which interpolates the two limits, of the degenerate Fermi liquid and non-degenerate correlated plasma. It appears that all nonmonotonic $m^*(n)$ dependencies for various temperatures collapse onto a single curve, when we choose $\alpha=0.4$, $\beta=1$, and $\gamma=1$  (see the inset in Fig.~\ref{Fig1}b). Some supporting reasonings from the plasma physics can be found in Supplementary Note 7, though the precision of our measurements is not too high to exclude other possible  $\tilde{r}_{\rm s}(n,T)$ functional forms.

{\bf Role of the in-plane magnetic field.}
In order to have a deeper insight into the  effective mass renormalization in the low density regime,  we  repeat the same measurements with the in-plane magnetic fields $B_\| =5.5$, and  9\,T,
 which produce Zeeman splitting $E_Z\sim 0.5$ and  1\,meV, respectively (see Fig.~\ref{Fig1}c).
 At low densities ($E_{\rm F}\lesssim T$, region $\mathbf{A}$ in Fig.~\ref{Fig1}c) the plasma is spin-polarized by $B_\|=9$\,T. Therefore, both, $S$ and $\partial S/\partial n$ at $B_\| = 9$\,T are expected to be less than the respective zero field values.

Region $\mathbf{A}$ is located in the vicinity of the critical density for sample Si-UW2 ($n_c\approx 8\times 10^{10}$cm$^{-2}$) and below it. If the free spins existed in the 2D system in the region $\mathbf{A}$, as the Mott-Wigner scenario of the 2D MIT predicts \cite{camjayi}, they would be fully polarized by the magnetic field $g\mu_{\rm B} B>T$ (i.e. at both $B_\| = 9$\,T and $B_\| = 5.5$\,T), and the entropy
would fall significantly, by $n_{\rm s} \ln 2$ (where $n_{\rm s}$ is the density of the free spins).
 However, in 5.5 Tesla field ($g\mu_{\rm B}B= 8$K$> T =3.2 $K), there is only a small ($\sim 10$\%) drop of the entropy per electron, which well corresponds to a partial polarization of itinerant electrons, as seen in Fig.\ref{Fig1}c.

To further show that the drop of entropy in parallel field is simply due to Zeeman polarization of the plasma, we extract the  effective mass for the Fermi gas [using Eq.~(4)  for $B=0$ as it was done above and Eq.~(8) from Supplementary note 8 for the Zeeman field] from our  $B=0$ and  9 Tesla $\partial S/\partial n$ data (see Fig.~2c).
The resulting mass is shown in Fig.~\ref{Fig1}c  with empty symbols. In region $\mathbf{A}$ (plasma regime), the effective masses for  $B=0$ and $B=9$\,T agree with each other to within 10\%, thus justifying our Fermi-plasma approach.

In region $\mathbf{C}$, for both $B=0$ and  $B=9$T-cases  the system is almost  Fermi-liquid,  where as explained above  $\partial S/ \partial n$ should be negative and  small. In the crossover (plasma-to-liquid) regime $\mathbf{B}$ the $\partial S/\partial n$  signal changes sign (Fig.~\ref{Fig1}c). For 9T-field the sign change happens at higher density than that for $B=0$; respectively,  the extracted mass at 9T exceeds essentially the one at zero field.
  Qualitatively, this is because  the nondegenerate minority subband produces the gaseous-like large contribution to the 9T data.  The crossover regime requires more detailed  consideration.

In region $\mathbf{C}$, the density of states and hence the entropy should be the same at 9\,T and at zero field.  This means that the total area under the $\partial S/\partial n$ curves in Fig.~\ref{Fig1}c should be the same. Indeed the hatched areas in Fig.~\ref{Fig1}c do compensate each other.

{\bf Integration of entropy in zero field.} Integration of the $\partial S/\partial n$ signal from zero density produces the entropy shown in Fig.\ref{Fig1}d. We neglect the low-density experimentally inaccessible domain $n<n_{\rm th}$ because it may lead only to a slight shift of the result, as shown by the dashed lines in Fig.~\ref{Fig1}d.
 Plotting the extracted entropy as a function of temperature, we see that for high electron density ($n =1.03 \cdot 10^{12}$\,cm$^{-2}$) it develops linearly-in-$T$ (see insert to Fig.~2d), extrapolating to zero at zero temperature, as it should be for a FL according to the 3rd law of thermodynamics.

For lower density ($n=3.9 \cdot 10^{11}$\, cm$^{-2}$) the $S(T)$ dependence does not extrapolate linearly to zero at $T=0$. It is possible that for this density the low-$T$ FL regime is not achieved yet, and  for even lower temperatures $S(T)$ should tend to zero with an increased slope, corresponding to the enhanced effective mass.

{\bf Role of disorder}
To further elucidate whether the mass growth originates from interelectron interactions or from disorder, we performed similar measurements with a more disordered sample Si8-9 whose mobility is lower by a factor of six
(see Fig.~\ref{compar} and Supplementary Note 9 where transport data are shown ).  We note that sample Si8-9 shows no signatures of the metal-to-insulator transition.
The threshold density in Si8-8 is much higher than in Si-UW2 (see inset in Fig.~\ref{compar}), therefore only the data above $n=1.5\times10^{11}$cm$^{-2}$ are available.
Firstly, as well as for the high-mobility sample SiUW-2, the signal ($\partial S/\partial n$) for Si8-9  exceeds considerably the  expected signal for the Fermi gas (dashed lines in Fig.~\ref{compar}). Secondly, within the density range where data are available for both samples, they are close to each other for all three temperatures.
The above facts signify that the effective mass enhancement is disorder-independent.

{\bf Discussion}

Interestingly, the explored  domain of low densities ($n<3 \cdot 10^{11}$cm$^{-2}$) and temperatures (2.5\,K$<T<5$\,K) belonging to the correlated plasma regime is exactly the regime of the strong ``metallicity'' and  the metal-insulator transition (MIT) for  2D electron systems in  high mobility Si-MOSFETs \cite{kravchenko, clarke-natphys_2008}.  The observation that the
domain of densities in the vicinity of the 2D MIT critical density falls into the strongly-interacting plasma regime
and can be parametrized with the temperature-dependent interaction parameter
is novel for MIT physics.
It
points  to the necessity of refining the existing theoretical approaches,
because
most of theoretical models \cite{camjayi,altmas, finkelstein_Science, spivak, dsh, meir} assume degenerate system.
We note that our findings do not exclude quantum phase transition or divergency of the effective mass at even lower temperatures.

 It is worth of noting here that everywhere we indirectly assume that our system is uniform down to the lowest densities. We can do so because we probe the 2D system on the (0.3-1)s time scale rather than the ps-scale as in transport measurements. For this reason, the regions (if any), poorly-conductive in transport, are well rechargeable in entropy measurements and the criteria of the inhomogeneity (that is a drop in the rechargeable area, i.e. in the capacitance) is shifted deep to the lower density side. In our case for high mobility Si samples we do not observe drop of the capacitance down to half the critical density (see the insert to Fig.~\ref{compar}). This result is similar to that for  n-GaAs in Ref.\cite{allison}) and therefore we  believe that our system is relatively uniform, in the above sense and in the explored elevated temperature range. In Supplementary note 10 we give numerical arguments that the potential nonuniformity is not essential for thermodynamic $\partial S/\partial n$ measurements unless the sample fails to recharge.

To summarize, in order to measure the equilibrium thermodynamic entropy of 2D electron systems we developed the capacitive technique,  whose  sensitivity enables measurements with $10^8$ electrons only. The key-feature of the
method consists in measuring the entropy per electron, $\partial S/\partial n$,  rather than the heat capacitance $c=TdS/dT$ as in the conventional ac-calorimetry.
We traced the entropy evolution in the wide ranges of electron density and temperature, from the  low-temperature FL regime to the regime of non-degenerate strongly correlated electron plasma.

We demonstrate that the plasma regime may be phenomenologically mapped to the ideal Fermi gas with a renormalized mass.
To characterize the renormalization of the effective mass we introduce an effective interaction parameter $\tilde{r}_{\rm s}$ which is {\it temperature and density dependent}. Our experimental findings are challenging for the theory of the strongly correlated 2D plasma.

{\Large \bf Methods}

{\bf Temperature modulation and measurements.}
In order to modulate the temperature and to avoid its possible gradients we used a copper cylindric sample holder of the diameter $d=$ 10-15 mm and the height $h=4$ mm with two $r= 100$\,Ohm twisted pair manganine heaters wound onto it. One heater was used to control the average temperature of the holder, the other was used for temperature modulation.

The sample and thermometer were glued to the bottom of the holder with heat-conductive varnish. As a thermometer to detect the holder temperature we used either Au:Fe-Cu thermocouple or RuO$_2$ chip resistor. The RuO$_2$ resistance was measured at elevated frequency ($\sim 1000$ Hz) and then its modulation at the frequency $f$ was detected with a separate lock-in amplifier.

{\bf Entropy per electron measurements.}
 The signal $\partial S/\partial n$ was detected on a double modulation frequency $2f$. For details concerning the choice of the frequency and explanation why the electrons are in thermal equilibrium with the sample we refer to the Supplementary Note 1.

In order to measure femtoampere modulation current a special care was taken to shield the sample from the heater, to ground the measurement circuit properly and to suppress vibrations. The scheme of the home-made amplifier (current-to-voltage converter) and the analysis of its operation are given in Supplementary Note 2.

{\bf Samples and characterization.} 
The parameters of the samples used are summarized in the Supplementary Table 1.
Electron density in the studied systems was determined by using either the low-field Hall effect or the magnetocapacitance dips in the quantum Hall Effect regime. Both methods give almost the same density value. See Supplementary Note 11 for details.
For GaAs gated heterojunctions, measurements and integration of $\partial S/\partial n$ in a wide range of densities were impossible because of
extremely large equilibration times and pA-level leakage in Schottky-gate at low densities (high negative gate voltages) .

{\bf Theoretical background.}
All required theoretical basis  is summarized in Supplementary Note 8 (2D systems in zero magnetic field) and Supplementary Note 4 (2D systems in finite magnetic field).

{\bf Acknowledgements}

We thank I. Gornyi, G. Minkov, D. Polyakov and M. Reznikov for discussions.
The work was funded by the Russian Science Foundation under the grant No. 14-12-000879.
The measurements were done using the equipment of the LPI shared facilities center.

\begin{widetext}
\clearpage
{\Large \bf Strongly correlated two-dimensional plasma explored from entropy measurements. \\SUPPLEMENTARY ONLINE MATERIALS.}
~\\

{\large \bf Supplementary Note 1. Temperature modulation}

The modulation heater was excited with a current $i= i_0 \cos(2\pi t f/2)$ at a frequency $f/2$. Correspondingly, the heating power $W$ was modulated at the double frequency $f$ (typically, $f=0.5$ Hz):
\begin{equation}
W = \frac{1}{2} i_0^2 r \Bigl [1+\cos(2\pi f t)\Bigr ]
\end{equation}

\begin{figure*}[H]
\begin{center}
\centerline{\psfig{figure=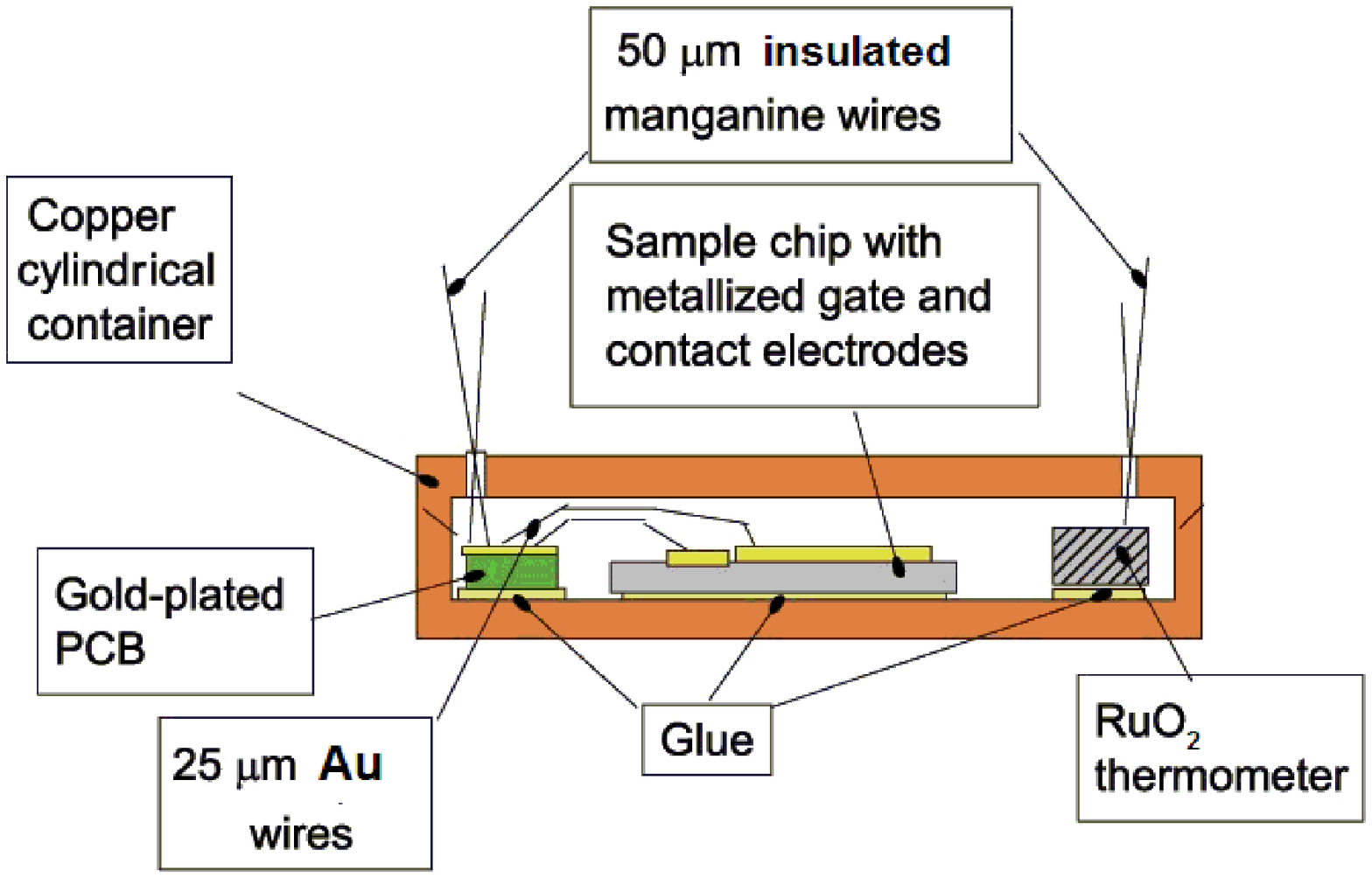, width=300pt}}
{{\bf Supplementary Figure 1. The schematics of the sample chamber.} Diameter of the can is 14 mm, height is 4 mm.}
\label{sampleholder}
\end{center}
\end{figure*}

The sample and the thermometers are placed in a cylindrical copper sample holder. The thermal equilibration time of the sample holder at 4.2 K  is short; it may be estimated  as $d^2 c \rho/\lambda \sim 10^{-3}$s where $\rho$ denotes the density,  $\lambda$ the thermal conductivity,  $c$  the specific heat, and  $d^2$ is the characteristic sample area across the heat flow. The equilibration time of the single-crystalline silicon chip is also small, $\sim 10^{-3}$\,s.The low equilibration times ensure the absence of the temperature gradients over the sample area.
When the sample, sample holder and thermometer are glued together and placed in a 4He-heat-exchanging gas (see Supplementary Fig. 1) the equilibration time increases being limited by heat conductance of the glue. The thermalization time of the RuO$_2$ thermometer is larger because of its ceramic body. The sample holder aims to achieve thermal equilibrium of all components inside it at a temperature modulation period. It is actively heated with the wire heater and is passively cooled via the heat sink to the cold bath. To minimize the heat flow directly from the sample and thermometer, they are thermally disconnected from the cold bath by the ~10 cm long manganin wires. We note that  a similar set-up is also used in the ac-calorimetry (see Ref. \cite{kraftmakher} for a review).

To evaluate the equilibration time experimentally we measured $\partial \mu/\partial T$ versus electron density at various modulation frequencies (see Supplementary Fig. 2). As one can see, at low frequencies (below 1 Hz), the signal is frequency-independent. It means that both the sample and thermometer have the same temperature  modulation $\Delta T$. As frequency increases, the signal starts to deviate from the initial value. It means that  $\Delta T$   for the sample and thermometer become different (i.e., thermal link to the sample is too weak). We therefore restricted our measurements to low frequencies ($f<1$Hz).

\begin{figure}[H]
\begin{center}
\centerline{\psfig{figure=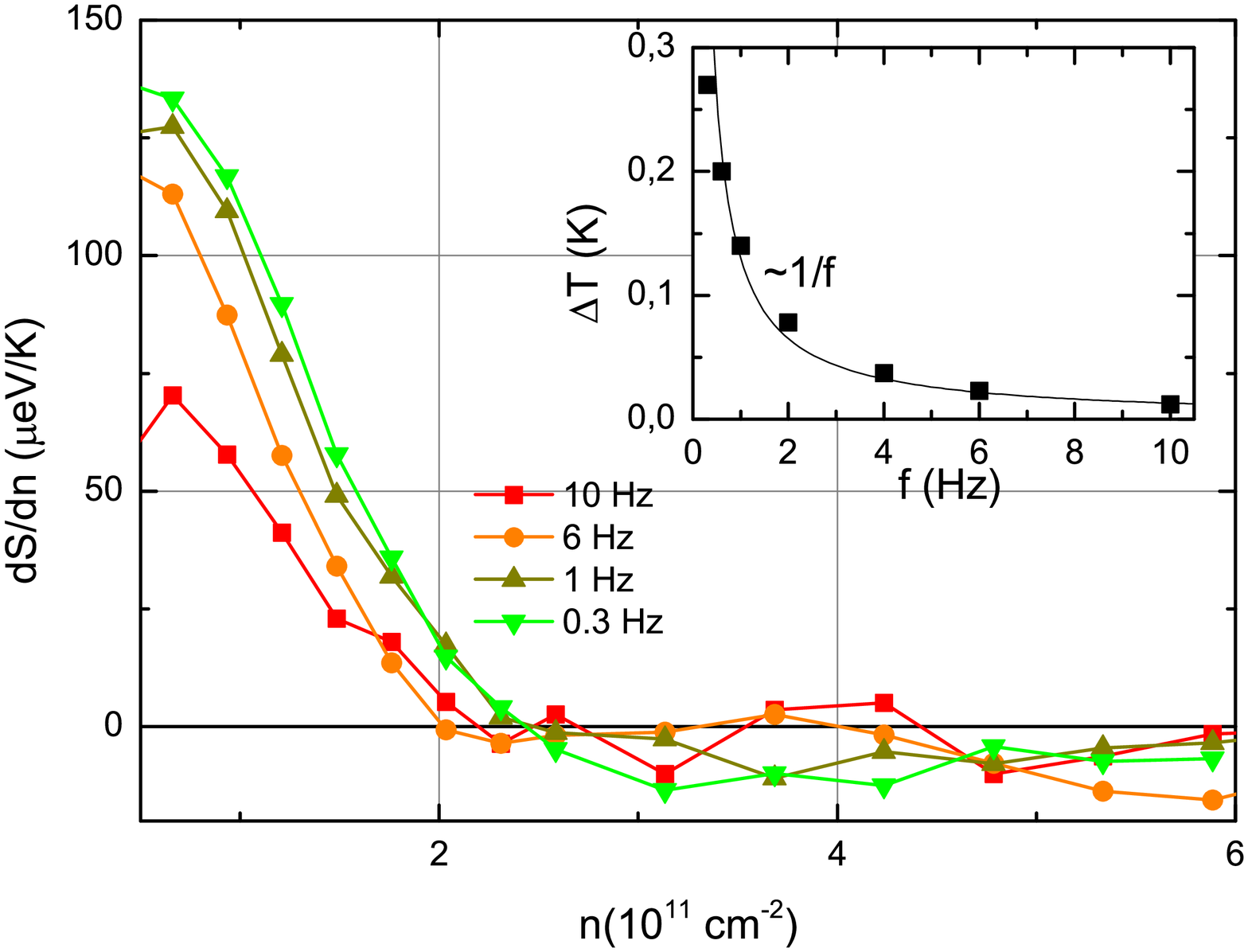, width=250pt}}
{{\bf Supplementary figure 2. Measured $\partial S/\partial n$ versus electron density for sample Si-UW2 at different frequencies $f$.} $T=2.6K$, $B=0$.}
\end{center}
\end{figure}

Provided the thermal equilibration time $\tau_0$ of the holder with the sample is much less than $1/f$, one can phenomenologically describe the system as a heat reservoir of the  capacitance $C$ connected to the cold bath with a weak link   $W_{\rm link}=\alpha (T-T_{\rm bath})$. Modulated heating causes: (i) increase of the temperature of the holder and (ii) its modulation. For the modulation one can write the following heat balance equation:
\begin{equation}
C \frac{d(\Delta T)}{dt}-\alpha\Delta T= \frac{1}{2} i_0^2r\cos(2\pi f t) .
\end{equation}
It's solution can be written as
\begin{equation}
\Delta T=\Delta T_0 \cos(2\pi f t+\phi) ,
\end{equation}
where
\begin{equation}
\Delta T_0= \frac{i_0^2 r}{2\sqrt{(2\pi f C)^2+\alpha^2}},\qquad
\tan \phi= \frac{2\pi f C}{\alpha} .
\end{equation}

In the high frequency limit,  $f > \alpha/C$ (several Hz in our case), the temperature modulation amplitude $\Delta T_0= i_0^2r/(4\pi f C)$ is inversely proportional to the frequency (see insert to Supplementary Fig. 2). The measured current (``signal'' for shortness) $2\pi f C_0\Delta T_0 \partial \mu/\partial T$ is thus frequency independent (see Eq.~(1) in the main text).

An important issue for low-temperature  measurements with 2D systems is why we are sure that electrons are in a thermal equilibrium
with the lattice. This problem is in fact crucial only at ultra low temperatures. For our experiments, at $T>2.5$\,K, there are solid arguments for the absence of electron overheating. The thermalization  of electron system with the lattice is very effective, its rate is $\sim \kappa/C_{el}\sim 3\times 10^7$s$^{-1}$ at 3K, where $\kappa\approx 2\times 10^{-5}$W/K cm$^2$ is the energy relaxation rate measured for similar Si-MOSFET samples in Ref. \cite{prus} and $C_{el}\approx 7\times10^{-13}$J/Kcm$^2$ is the specific heat of 2D electrons in (100)Si-MOSFETs at 3K. Therefore, the  electron temperature may differ from the holder temperature only due to potential overheating by external RF fields. In a separate experiments we checked  that in a similar  environment there are no signatures of the overheating (the phase breaking time develops as $1/T$ without saturation, and Shubnikov-de Haas amplitude grows as temperature decreases down to 1.5\,K). This is not surprising because we have filtered the side high-frequency signals in our measurements set-up.

~\\
{\large \bf Supplementary note 2. Measurement setup and samples studied.}
~\\
For most of measurements the power modulation frequency, $f=0.624$\,Hz, was chosen  as a compromise between the noise level, possibility to achieve low temperatures and possibility to access the low electron densities deep in the insulating regime. In our experiment the lowest available base temperature is 2 K without modulation heating. With increase of the amplitude of temperature modulation  the average temperature grows. The amplitude of temperature modulation determines the sensitivity of our measurements. We  chose the modulation amplitude equal to 0.05\,K for which  the lowest temperature was equal to 2.4\,K. To access the low electron densities deep in the insulating regime the modulation frequency is restricted by the inverse  recharging time,  $f<1/(2\pi R C_0)$, where $R$ is the resistance of 2DEG and the contact, and $C_0$ is the geometrical capacitance.

From the electrical point of view, temperature in our measurement is just an external parameter; it does not differ from, e.g., magnetic field $B$. The procedure to measure $\partial \mu/\partial B$  in the insulating regime ($2\pi f RC \sim 1$) is justified in Ref. \cite{jetpl}. One needs to measure a complex capacitance at the same frequency as the signal, to account for the phase shift. The same procedure is applicable here and allows us to perform measurements deep in the insulating regime ($R\sim 1$GOhm, $n\sim 3\cdot 10^{11}$ cm$^{-2}$).

\begin{figure}[H]
\begin{center}
\centerline{\psfig{figure=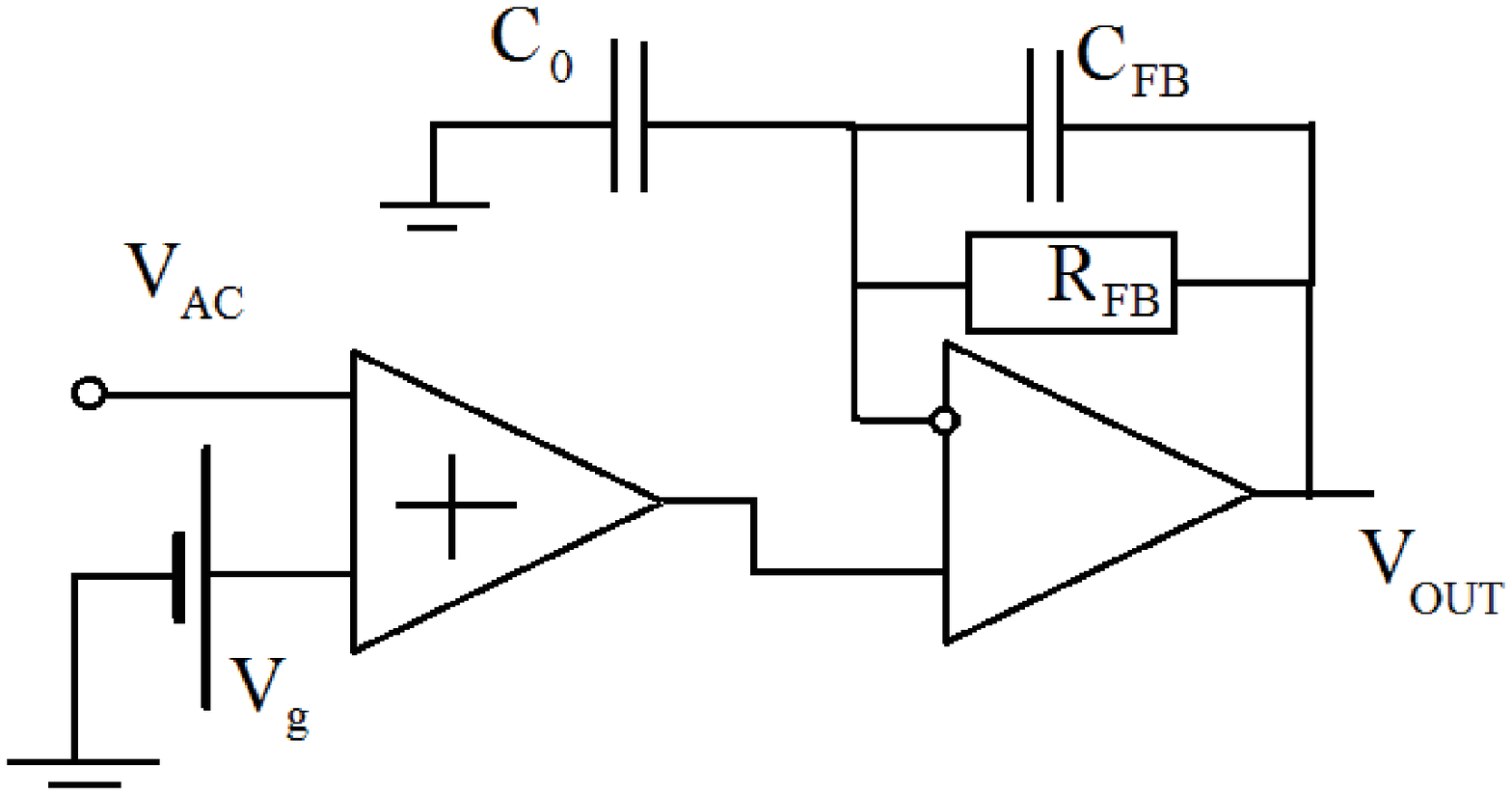, width=300pt}}
{{\bf Supplementary Figure 3. The diagram of the measurement circuit.} $C_0$ is the sample with grounded 2DEG, $V_g$ is an external voltage source. $R_{FB}\sim 10$ GOhm and $C_{FB}\sim 10$ pF form the feedback impedance.}
\end{center}
\end{figure}

The diagram of our measurement setup is shown in Supplementary Fig. 3.   The sample is shown as capacitor $C_0$. When  an external  parameter is modulated, the ac current $I_{AC}$ flows through the feedback (FB) of the current amplifier and causes ac voltage at the output $V_{OUT}=-I_{AC}Z_{FB}$, where $Z_{FB}=R_{FB}/(1+i 2\pi f C_{FB} R_{FB})$. We use AD 820 operational amplifier with input impedance $10^{13}$ Ohm$||0.5$ pF. The output voltage is measured with a lock-in amplifier. The gate voltage is set by the external voltage source through the summing amplifier.
 The  sample capacitance  value $C_0$ can be measured in situ, using the $V_{AC}$ input. An ac voltage at the input causes the current $I_{AC}=-2\pi f C_0 V_{AC}$ through the feedback.

To separate the physics of interactions and disorder we selected for these studies several samples: high-mobility Si-MOSFETs Si-UW1 and Si-UW2, low-mobility Si-MOSFET Si8-9 and Schottky-gated GaAs/AlGaAs heterojunction. The relevant sample parameters are summarized in the Supplementary Table 1.
\begin{table}[H]
{Supplementary table 1. Samples' properties\\}
\begin{tabular}[t]{|c|c|c|c|c|c|}
\hline
Sample & Density range, &$E_F$ range, &Capacitance, &Area,  &Peak mobility,\\
&10$^{11}$ cm$^{-2}$&meV&pF&mm$^2$&m$^2/$Vs\\
\hline
SiUW1 & 0.3-12 &0.2-7.5& 700 &4&3\\
\hline

SiUW2 & 0.3-12 &0.2-7.5& 680 &4&3\\
\hline
Si8-9 & 1.5-12 &0.9-7.5& 630 &4&0.5\\
\hline
GaAs1 & 0.4-5 &1.7-20& 1100 &5&20\\
\hline
\end{tabular}
\end{table}

~\\
{\large \bf Supplementary note 3.Testing the technique in quantizing magnetic fields.}
~\\
In order to obtain a quantitative information from magnetooscillations of  $\partial S/\partial n$, we fit the data according to Eqs.~(\ref{eq:12}) and (\ref{eq:12:Z}). The fit  was performed using three adjustable parameters, the level broadening $\Gamma$, effective mass and $g$-factor. In fact, the effective mass and $g$-factor are very close to their band values. In particular, for GaAs1 sample one can neglect Zeeman effects below 3 Tesla. The examples of data fit are shown in Figs.~1b and 1d of the main text.
\begin{figure}[H]

\begin{center}
\centerline{\psfig{figure=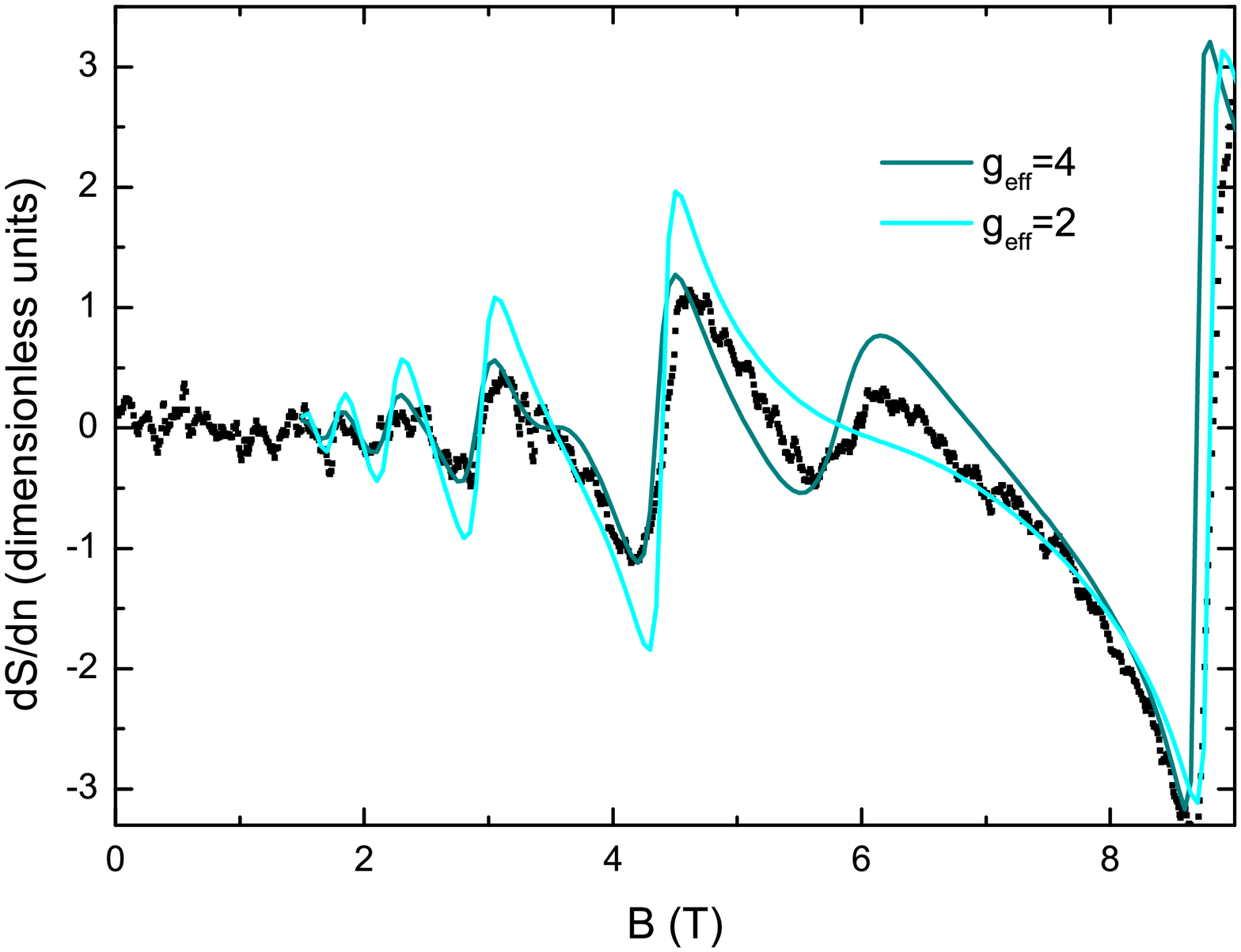, width=200pt}}
{{\bf Supplementary Figure 4. Magnetooscilations of the entropy per electron.} Black dots show the $\partial S/\partial n(B)$ dependence for sample SiUW1 at $n=8.5 10^{11}$cm$^{-2}$ for $T=3$K (dots). Light cyan line is a fit using Eq.\ref{eq:12:Z} with $m=m_b$, $g=2$,$\Gamma=0.3$meV; dark cyan line is the fit with $m=m_b$, $g=4$,$\Gamma=0.3$meV.}
\end{center}
\end{figure}

In higher field and at low temperatures the interlevel exchange interaction increases the spin and cyclotron gaps (precursors of quantum Hall ferromagnetism \cite{Girvin}), and opens the valley gaps
\cite{pudalov-valley_1985} in Si-based structures. This makes a simple noninteracting theory inapplicable. In particular, features related with the valley gaps are clearly seen at $\nu=1, 3$ in Figs. 1i and 1j of the main text. Quantitatively, in order to account for the gap increase we may use enhanced g-factor. Supplementary Figure 4 illustrates two  examples of fitting the same $\partial S/\partial n (B)$ data for Si-UW1 sample with $g=2$ (bare value) and  with enhanced $g_L=4$. The latter fit allows to resolve the $\nu=6$ Zeeman feature. We thus show that our data reproduce reliably the previous results. Detailed study of the oscillatory phenomena by means of entropy is beyond the scope of the present work and will be published elsewhere.

~\\
{\large \bf Supplementary Note 4.Thermodynamics in a perpendicular magnetic field}
~\\

\subsubsection{2D noninteracting electrons}

In the presence of the perpendicular magnetic field $B$ the thermodynamic potential is given as (we disregard the Zeeman splitting for a moment)
\begin{equation}
\Omega(T,\mu,B) = -T \int d\varepsilon\, D(\varepsilon) \ln \left [ 1+e^{(\mu-\varepsilon)/T} \right ] ,
\label{eq:Omega:B}
\end{equation}
where the density of states
\begin{equation}
D(\varepsilon) = g_{v,s} \frac{m\omega_c}{2\pi} \sum_{n=0}^\infty  \mathcal{W}\bigl (\varepsilon-\omega_c(n+1/2)\bigr ) .
\end{equation}
Here we introduce the function $\mathcal{W}(\varepsilon)$ to describe the broadening of a Landau level by disorder. It satisfies the normalization condition:
$\int d\varepsilon\, \mathcal{W}(\varepsilon) = 1$.
The typical width of a Landau level can be estimated as
$\Gamma \sim \left [  \int d\varepsilon\,  \mathcal{W}^{\prime\prime}(\varepsilon) \right ]^{-1/2}$.
As usual, by means of the Poisson resummation formula, for $\mu\gg \omega_c, T, \Gamma$ we find the Lifshitz-Kosevich-type expression \cite{LK}:
\begin{equation}
\Omega(T,\mu,B) = \Omega(T,\mu,B=0) + \frac{g_{v,s} m\omega_c^2}{4\pi^3} \sum_{k=1}^{\infty}
\frac{(-1)^{k-1}}{k^2} \left [ 1 - \mathcal{A}_k \frac{(2\pi^2 T k/\omega_c)}{\sinh (2\pi^2 T k/\omega_c)} \cos \frac{2\pi \mu k}{\omega_c}\right ] \, .
\label{eq:Om:B}
\end{equation}
Here $\Omega(T,\mu,B=0)$ denotes the zero-field thermodynamics potential and the parameter $\mathcal{A}_k$ characterizes the Landau level broadening:
\begin{equation}
\mathcal{A}_k  = \int d\varepsilon\, \mathcal{W}(\varepsilon) \exp\left ( \frac{2\pi i\varepsilon k}{\omega_c} \right )\,.
\end{equation}
In the absence of disorder $\mathcal{W}(\varepsilon) = \delta(\varepsilon)$ and $A_k=1$. For commonly used models of the disorder Landau level broadening the functions $\mathcal{W}(\varepsilon)$ and the Fourier coefficients $\mathcal{A}_k$ are summarized in Supplementary Table 2.
\begin{table}
{Supplementary table 2. The function $\mathcal{W}(\varepsilon)$ for different models of Landau level broadening. Here $J_1(x)$ denotes the Bessel function.}
\\
\begin{tabular}{lc|clc|cl}
{\rm Lorentzian} & & & $\mathcal{W}(\varepsilon) = \frac{1}{\pi} \frac{\Gamma}{\varepsilon^2+\Gamma^2}$ & & & $\mathcal{A}_k= e^{-2\pi \Gamma |k|/\omega_c}$\\
{\rm Gaussian} & & & $\mathcal{W}(\varepsilon) = \frac{1}{\Gamma\sqrt{\pi}} e^{-\varepsilon^2/\Gamma^2}$ & & &
$\mathcal{A}_k= e^{-\pi^2 \Gamma^2 k^2/\omega^2_c}$ \\
{\rm Semicircle} & & & $\mathcal{W}(\varepsilon) =  \frac{2}{\pi \Gamma} \sqrt{1-\varepsilon^2/\Gamma^2}$ & & &  $\mathcal{A}_k=\frac{\omega_c}{\pi \Gamma k} J_1\left (\frac{2\pi \Gamma k}{\omega_c}\right )$
\end{tabular}
\label{def:Tab1}
\end{table}

We mention that as follows from Eq. \eqref{eq:Omega:B} in the limit of small temperature $T\ll \omega_c, \Gamma, \mu$, the entropy is given by the result \eqref{eq:ss}. In the case of the Lorentzian Landau level broadening the density of states at the Fermi energy becomes
\begin{equation}
D(E_F) = \frac{g_{v,s} m}{2\pi} \frac{(1/2) \sinh (2\pi \Gamma/\omega_c)}{\sinh^2(\pi\Gamma/\omega_c)+\cos^2(\pi E_F/\omega_c)}
 \,  .
 \label{eq:S:T0:H}
\end{equation}
In the limit $\Gamma\ll \omega_c$ Eq. \eqref{eq:S:T0:H} implies that the entropy is almost zero for the chemical potential outside a Landau level, $E_F \neq \omega_c(N+1/2)$: $S \sim  \pi^2 g_{v,s} m T \Gamma/(6\omega_c)\ll \pi g_{v,s} m T/6$. In case of the chemical potential inside a Landau level, $|E_F -\omega_c(N+1/2)|\lesssim \Gamma$, the entropy is large: $S =  g_{v,s} m T \omega_c/(6\Gamma)  \gg \pi g_{v,s} m T/6$.

Equation \eqref{eq:Om:B} results in the following dependence of the electron density on the chemical potential, temperature and magnetic field
\begin{equation}
n = \frac{g_{v,s} m \mu}{2\pi}  -  \frac{g_{v,s} m\omega_c}{2\pi} \sum_{k=1}^{\infty}
\frac{(-1)^{k-1}}{\pi k} \mathcal{A}_k \frac{(2\pi^2 T k/\omega_c)}{\sinh (2\pi^2 T k/\omega_c)} \sin \frac{2\pi \mu k}{\omega_c}\,  .\label{eq:n}
\end{equation}
Using the following thermodynamic identity (see e.g. \cite{LL5})
\begin{equation}
\left ( \frac{\partial n}{\partial T} \right )_\mu = - \left ( \frac{\partial \mu}{\partial T} \right )_n  \left ( \frac{\partial n}{\partial \mu}\right )_T =\left ( \frac{\partial S}{\partial n} \right )_T  \left ( \frac{\partial n}{\partial \mu}\right )_T  \, ,
\end{equation}
we find that
\begin{align}
\left ( \frac{\partial S}{\partial n} \right )_T & = -\frac{\omega_c}{T} \sum_{k=1}^{\infty}
\frac{(-1)^{k-1}}{\pi k}  \mathcal{A}_k \frac{(2\pi^2 T k/\omega_c)}{\sinh (2\pi^2 T k/\omega_c)}\left [ 1 -\frac{2\pi^2 T k}{\omega_c} \coth \frac{2\pi^2 T k}{\omega_c} \right ]   \sin \frac{2\pi \mu k}{\omega_c} \notag \\
& \times \left [ 1 -  2 \sum_{k=1}^{\infty}
(-1)^{k-1} \mathcal{A}_k \frac{(2\pi^2 T k/\omega_c)}{\sinh (2\pi^2 T k/\omega_c)} \cos \frac{2\pi \mu k}{\omega_c}
\right ]^{-1}
 \, .
\label{eq:12}
\end{align}
Equations \eqref{eq:n} and \eqref{eq:12} determine the dependence of the derivative $\left ( {\partial S}/{\partial n} \right )_T$ on the electron density $n$, temperature and magnetic field.
In the case of weak oscillation, $A_k\ll 1$ ($\Gamma\gg \omega_c$), we can simplify Eq. \eqref{eq:12} as
\begin{align}
\left ( \frac{\partial S}{\partial n} \right )_T = -
  \frac{2\pi  \mathcal{A}_1}{\sinh (2\pi^2  T/\omega_c)}\left [ 1 -\frac{2\pi^2 T}{\omega_c} \coth \frac{2\pi^2 T }{\omega_c} \right ]   \sin \frac{4\pi^2 n}{g_{v,s} m \omega_c}
 \, .
\label{eq:12:s}
\end{align}

In the presence of the weak Zeeman splitting, $Z \ll E_F$ the partition function can be found by means of following substitution:
\begin{equation}
\Omega(T,\mu,B) \to \frac{1}{2}\Omega(T,\mu+Z,B)+\frac{1}{2}\Omega(T,\mu-Z,B) \, .
\end{equation}
Then for $\mu\pm Z \gg \omega_c, T, \Gamma$ we find
\begin{equation}
\Omega(T,\mu,B) = \frac{1}{2}\sum_{\sigma=\pm } \Omega(T,\mu+\sigma Z ,B=0)+ \frac{g_{v,s} m\omega_c^2}{4\pi^3}  \sum_{k=1}^{\infty}
\frac{(-1)^{k-1}}{k^2} \left [ 1 - \mathcal{A}_k \frac{(2\pi^2 T k/\omega_c)}{\sinh (2\pi^2 T k/\omega_c)} \cos \frac{2\pi \mu k}{\omega_c}\cos \frac{2\pi Z k}{\omega_c}\right ] \, .
\end{equation}
Hence the electron density is given as
\begin{equation}
n(T,\mu,H) = \frac{g m \mu}{2\pi}  -  \frac{g m\omega_c}{2\pi}\sum_{k=1}^{\infty}
\frac{(-1)^{k-1}}{\pi k} A_k \frac{\frac{2\pi^2 T k}{\omega_c}}{\sinh \frac{2\pi^2 T k}{\omega_c}} \sin \frac{2\pi \mu k}{\omega_c}  \cos \frac{2\pi Z k}{\omega_c}\, , \label{eq:n1}
\end{equation}
and the derivative of the entropy becomes as follows
\begin{align}
\left ( \frac{\partial S}{\partial n} \right )_T & = -\frac{\omega_c}{T} \sum_{k=1}^{\infty}
\frac{(-1)^{k-1}}{\pi k}  \mathcal{A}_k \frac{(2\pi^2 T k/\omega_c)}{\sinh (2\pi^2 T k/\omega_c)}\left [ 1 -\frac{2\pi^2 T k}{\omega_c} \coth \frac{2\pi^2 T k}{\omega_c} \right ]   \sin \frac{2\pi \mu k}{\omega_c}  \cos \frac{2\pi Z k}{\omega_c}\notag \\
& \times \left [ 1 -  2 \sum_{k=1}^{\infty}
(-1)^{k-1} \mathcal{A}_k \frac{(2\pi^2 T k/\omega_c)}{\sinh (2\pi^2 T k/\omega_c)} \cos \frac{2\pi \mu k}{\omega_c} \cos \frac{2\pi Z k}{\omega_c}
\right ]^{-1}
 \, .
\label{eq:12:Z}
\end{align}

~\\

\subsubsection{Interacting 2D electron system}
~\\

For 3D electron system the Lifshitz-Kosevich expression for magnetooscillations of the thermodynamic potential obtained in the absence of interaction can be used for the case of interaction provided one takes into account standard Fermi-liquid renormalization of  the quasiparticle spectrum \cite{L, BG}. For two-dimensional electron system, it is not the case, in general \cite{CS}. For classically weak perpendicular magnetic field, modification of the Lifshitz-Kosevich-type expression for a 2D electron interacting disordered system has been studied in Refs. \cite{MMR,AGM}.

As follows from the results of Ref. \cite{AGM}, for classically weak magnetic field, $\Gamma\gg \omega_c$, [in this case the Landau levels are almost completely broadened and $\Gamma=1/(2\tau)$], the derivative $\partial S/\partial n$ has small harmonic oscillations with the magnetic field:
\begin{align}
\left ( \frac{\partial S}{\partial n} \right )_T  = \frac{4\pi^3 T}{\omega^*_c}
 e^{-2\pi^2 T/\omega^*_c} e^{- \pi /(\omega^*_c\tau_*)}e^{{\mathcal B}(T)}   \sin \frac{2\pi n}{n_L}
 \, .
\label{eq:12:ss}
\end{align}
Here $ n_L = g_{v,s} e B/c$ is the Landau level degeneracy, and $\omega_c^*=e B/(m^* c)$. Apart from the standard Fermi-liquid renormalization of
the effective mass, $m\to m^*$ and the Dingle temperature $1/\tau \to 1/\tau_*$, there is additional temperature dependence of the amplitude of the oscillations due to combined effect of interaction and disorder which is described
by the $\mathcal{B}(T)$. In the ballistic regime, $4\pi T \tau \gg 1$, and for Coulomb type scattering the function ${\mathcal B}(T)$ is given as \cite{AGM}:
\begin{align}
{\mathcal B}(T) = \frac{\pi T}{\omega_c \sigma}  \left ( 1+(g_{v,s}^2 -1) \frac{F_0^\sigma}{1+F_0^\sigma}\right ) \ln \frac{2g_{v,s} m e^2 v_F}{\overline{\kappa }T}.
\end{align}
In the diffusive regime, $4\pi T \tau \ll 1$, the expression for ${\mathcal B}(T)$ can be found in Ref. \cite{AGM}.
~\\

{\large \bf Supplementary Note 5. Comparing with the ac calorimetry}
~\\

Strictly speaking, the method described above allows us to measure $\partial S/\partial n$, whereas the ac calorimetry measures specific heat $T\partial S/\partial T$. Both methods, however, allow one to evaluate the entropy change. In the quantizing regime, when the entropy oscillates one can integrate both $\partial S/\partial n$ and $\partial S/\partial T$, and to calculate the respective changes in entropy between its neighboring maximum and minimum. We compare the smallest oscillation one may resolve with each particular method.

The ac calorimetry measurements in a sample with density  $8.8\cdot 10^{11}$\, cm$^{-2}$, mobility $10$\,m$^2/$Vs (both comparable to our  GaAs1 sample) and in a similar temperature range (1.7- 4.6 K) were done with a stack  of 75 quantum wells, with 50 times larger total area  \cite{wang}. The smallest oscillation was observed at $\nu=12$ in 3 Tesla.

In our measurements with  GaAs1 sample  (Fig 1 in the main text) at 2.5 K we observed huge oscillations at the same field of 3 Tesla.  The oscillations were clearly observed down to 1\,Tesla,   where they had   a factor of 50 lower amplitude.  This implies that our technique has $\sim50\times50=2500$ times  better sensitivity per unit area than the  ac-calorimetry.  Remarkably, the  $(\partial S/\partial n)$ value we measure is of purely electronic origin and has no contribution from the lattice specific heat.

~\\

{\large \bf Supplementary note 6. Fermi-liquid regime}
~\\
\begin{figure}[H]
\begin{center}
\centerline{\psfig{figure=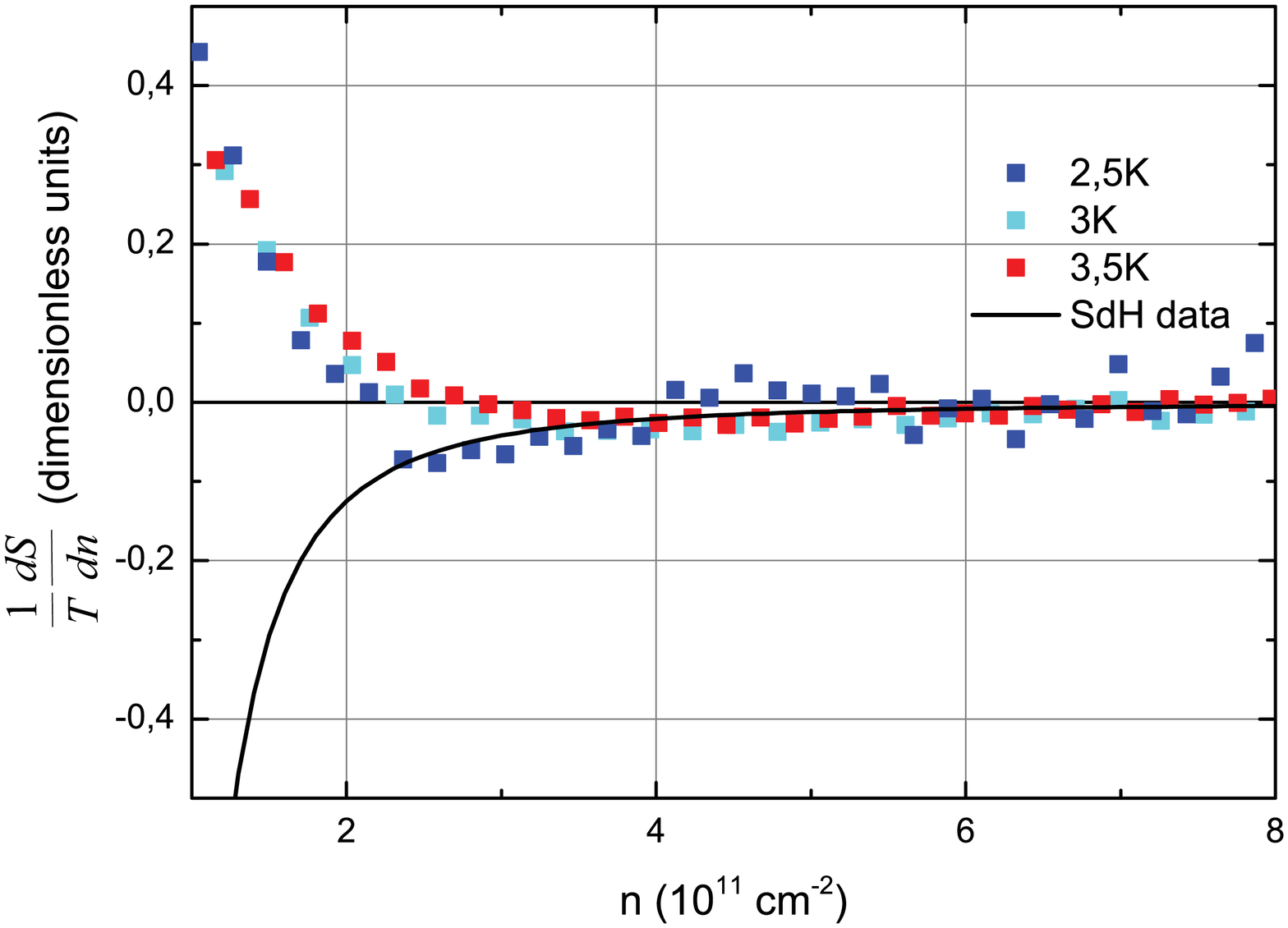, width=200pt}}
{{\bf Supplementary Figure 5. The lowest temperature negative $\partial S/\partial n$ signal}(normalized by temperature).}
\end{center}
\end{figure}

At the lowest temperatures in the Fermi-liquid regime the entropy per electron is negative, $\partial S/\partial n<0$, due to renormalization of the effective mass (see Ref.\cite{ando_review} and Supplementary Note 8). Using the definition of density of states:
\begin{equation}
D^*=g_{v,s}m^*/2\pi \hbar^2
\end{equation}
and definition of entropy in degenerate Fermi system $S=\pi^2 T D^*/3$ (here $m^*$ is the interaction renormalized mass, and we set $k_B=1$ throughout the Supplementary notes), we may compare the  lowest temperature negative $\partial S/\partial n<0$ data  with that calculated with the  effective mass independently measured from Shubnikov-de-Haas effect~\cite{PudFL}. The comparison in  Supplementary Fig. 5 shows a reasonable qualitative agreement.

~\\
{\large \bf Supplementary note 7.Degenerate and non-degenerate Fermi-liquid regime.}
~\\

In low temperature Fermi-liquid regime the strength of interaction is parameterized by the ratio of typical interaction and kinetic energies, $r_s=1/\sqrt{\pi n a_B^2} \sim U/E_F$. Here $U=e^2 \sqrt{n}/\overline{\kappa}$,
 the average dielectric constant at the interface $\overline{\kappa} =7.7$ \cite{ando_review},  and the Fermi energy $E_F= \hbar^2 k_F^2/2m_b$. At high temperatures, $T\gg E_F$, the kinetic energy is determined by the temperature. Therefore, one might expect that the effective interaction parameter $\tilde{r}_s$  should depend not only on electron density, but also on temperature. In the case of very high temperatures $T\gg U \gg E_F$, it is natural to choose $\tilde{r}_s$ to be proportional to the ratio $U/T$. At the intermediate temperatures $U\gg T\gg E_F$ there are other energy scales related with the interaction: i) the plasmon frequency at the Fermi wave vector $k_F$, $\omega_p(k_F) \sim \sqrt{E_F U}$; ii) the electron-electron collision rate which is of the order of $\omega_p(k_D) \sim U \sqrt{E_F/T}$ where $k_D \sim k_F \sqrt{U/T}$ denotes the 2D Debye wave vector \cite{PRA1,PRA2}. Therefore, we introduce  phenomenologically a generalized    $\tilde{r}_s(n,T)$ dependence  that interpolates the two known limits: $\tilde{r}_s=[\pi a_B^2 n+\alpha T^{\gamma+\beta}/E_F^\gamma U^\beta]^{-1/2}$.
Theoretically, we expect three different possibilities for the parameter $\tilde{r}_s$ at $T \gg E_F$:
i) $\tilde{r}_s$ is determined by the ratio $U/T$, i.e.,  $\gamma=0$ and $\beta=2$;
ii) $\tilde{r}_s$ is determined by the ratio $\omega_p(k_F)/T$, i.e.,  $\gamma=1$ and $\beta=1$;
iii) $\tilde{r}_s$ is determined by the ratio $\omega_p(k_D)/T$, i.e., $\gamma=1$ and $\beta=2$.  Although all three definitions might be formally possible, the last one, obviously, does not fit the highest temperature limit $T\gg U > E_F$.

\begin{figure}[H]
\begin{center}
\centerline{\psfig{figure=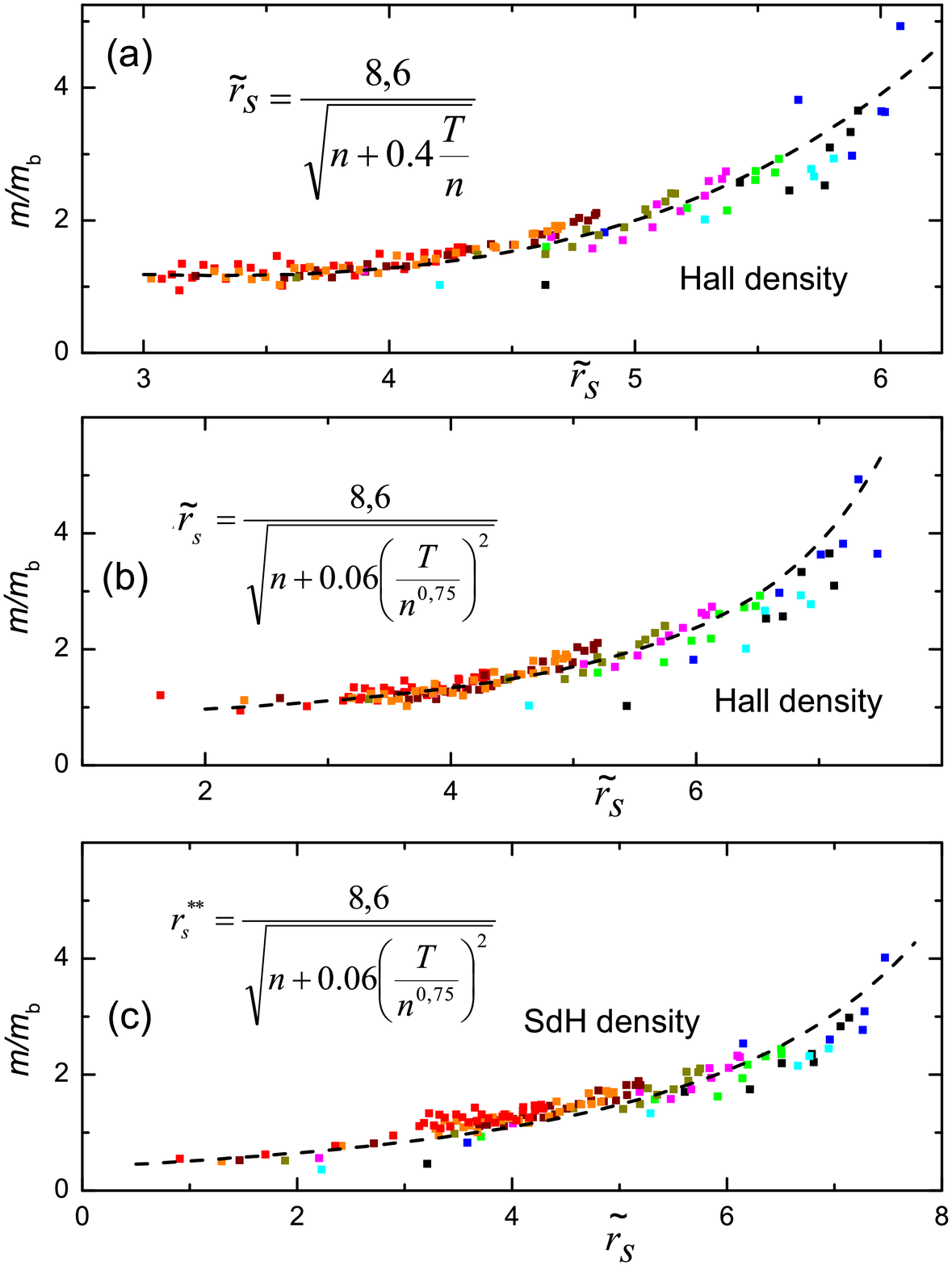, width=300pt}}
{{\bf Supplementary Figure 6. Effective mass versus effective interaction parameter $\tilde{r}_s$.} (a,b) The $\tilde{r}_s$ values were determined in various ways (the formulae are shown) using the Hall density. (c) The same as (b) using the density determined from SdH effect.}
\end{center}
\end{figure}

Experimentally our $\partial S/\partial n$ data are  somewhat scattered  and collected in a limited temperature and density range. The value of the effective mass relies also on measurement of the electron density which can be defined in two different manners giving different results. If we trust the Hall density, the best fit of experimental data is achieved for  $\alpha=0.4$, $\beta=0$ and $\gamma=1$ (see Supplementary Fig. 6a ). However, we cannot find for this fit a physical interpretation. Among possibilities described above the best fit is achieved for the case (ii) with  $\alpha= 0.4$ (see Supplementary Fig. 6b).
Changing the definition of density (``Hall density'' versus ``QHE density'') modifies slightly the optimal value of $\alpha$ and does not  improve considerably the agreement between the fit and the data  (see Supplementary Fig. 6c ).

~\\

{\large \bf Supplementary note 8. Thermodynamics in zero perpendicular magnetic field.}
~\\

\subsubsection{2D Ideal Fermi gas}
~\\

The thermodynamic potential per unit area of the ideal Fermi gas is given as (see e.g. \cite{LL5})
\begin{gather}
\Omega = - T \int d\varepsilon\, D(\varepsilon)\, \ln \left [ 1+e^{(\mu-\varepsilon)/T}\right ] . \label{eq:nS}
\end{gather}
Here $D(\varepsilon)$ is the single-particle density of states. Provided $D(\varepsilon)=g_{v,s} m/(2\pi)\theta(\varepsilon)$, where  $g_{v,s}$ takes into account possible spin and valley degeneracy,   $\theta(x)$ denotes  the Heaviside step function,  and we set $\hbar=1$  from here onwards,  we find
 the following temperature dependence of the chemical potential for a fixed electron density:
\begin{equation}
\mu = T \ln \left [e^{E_F/T} -1 \right ] .
\end{equation}
Here $E_F = 2\pi n/(g_{v,s} m)$ denotes the Fermi energy.
Hence the derivative of the entropy with respect to the electron density is as follows (see Eq. (4) in the main text)
\begin{equation}
\left ( \frac{\partial S}{\partial n} \right )_T  \equiv - \left ( \frac{\partial \mu}{\partial T} \right )_n  = \frac{E_F/T}{e^{E_F/T}-1}-\ln \left [1-e^{-E_F/T}  \right ] \, .
\end{equation}

In the presence of a parallel magnetic field $B$, the thermodynamic potential can be written as
\begin{equation}
\Omega(T,\mu,B) = \frac{1}{2}\Omega(T,\mu+Z)+\frac{1}{2}\Omega(T,\mu-Z) \, ,
\end{equation}
with $\Omega(T,\mu)$ given by Eq. \eqref{eq:nS}. Here $Z=g_L \omega_c/4$ stands for the Zeeman splitting ($\omega_c=eB/(mc)$).  Hence we obtain the following chemical potential at a fixed total electron density:
\begin{equation}
\mu = T \ln \left [\left (\sinh^2(Z/T)+e^{2E_F/T}\right )^{1/2} -\cosh(Z/T) \right ] .
\end{equation}
\begin{equation}
\mu = T \ln \left [\sqrt{\sinh^2\frac{Z}{T}+e^{2E_F/T}} -\cosh\frac{Z}{T} \right ] .
\end{equation}

Therefore the derivative of the entropy with respect to the electron density becomes
\begin{gather}
\left ( \frac{\partial S}{\partial n} \right )_T  \equiv - \left ( \frac{\partial \mu}{\partial T} \right )_n  =
 \frac{E_F/T}{1-e^{-2E_F/T}} \left [1+ \frac{\cosh(Z/T)}{\sqrt{\sinh^2(Z/T)+e^{2E_F/T}}} \right ]
 - \frac{(Z/T)\sinh(Z/T)}{\sqrt{\sinh^2(Z/T)+e^{2E_F/T}}} \notag \\  -  \ln \left [\sqrt{\sinh^2(Z/T)+e^{2E_F/T}} -\cosh(Z/T) \right ]
\, .
\end{gather}

~\\

\subsubsection{2D Fermi liquid}
~\\

The Fermi liquid concept is applicable to the low temperature regime $T\ll E_F$ only. The well-known low-$T$ expression for the specific heat (see e.g. \cite{LL9})
\begin{equation}
C(T) = \frac{\pi^2 D^*(E_F) T}{3} \, , \qquad T\ll E_F \, .
\label{eq:cc}
\end{equation}
implies the following expression for the entropy
\begin{equation}
S = \frac{\pi^2 D*(E_F) T}{3} \, , \qquad T\ll E_F \, .
\label{eq:ss}
\end{equation}
Here $D*(E_F) = g_{v,s} m^*/(2\pi)$ is the quasiparticle density of states at the Fermi level, the effective mass $m^*=m(1+F_1^\rho)$ depends on the electron density $n$ via the Landau coupling constant  $F_1^\rho$. Hence, we find
\begin{equation}
 \left (\frac{\partial S}{\partial n}\right )_T =
 \frac{\pi g_{v,s} T}{6}\frac{d m^*}{dn}  \, , \qquad T\ll E_F \, .
 \label{eq:sc}
\end{equation}

In fact, as well-known, there is non-analytic correction of the order $T^2$  to the lowest order Fermi-liquid result \eqref{eq:cc}. \cite{CB,MC,MPE} This implies that
 \begin{equation}
\left (\frac{\partial S}{\partial n}\right )_T   =
\frac{\pi g_{v,s} T}{6}\frac{dm^*}{dn} + \beta(n) T^2 \, , \qquad T\ll E_F \, .
 \label{eq:Sn:ff}
\end{equation}
In the limit of large electron density, $r_s\ll 1$
it is known \cite{CB} that
$
\beta(n) = 3 \zeta(3) g_{v,s} m^2/(16\pi^2 n^2)$. Here $\zeta(x)$ denotes the Riemann zeta function. In the limit of strong interaction, $r_s\gtrsim 1$, the function
$\beta(n)$
is determined by the quasiparticle backscattering amplitudes at the Fermi surface for charge and spin channels \cite{MC,MPE}.
We note here that the above formulae are given for the review purpose only. We did not observe signatures of the nonanalytic corrections in our experiments.

In the presence of weak short-ranged disorder, $E_F\tau\gg 1$, where $\tau$ stands for the elastic transport time, the function
 $\beta(n)$ can be written in terms of Landau interaction parameters in charge ($F_0^\rho$) and spin/isospin ($F_0^\sigma$) channels\cite{CA}:
\begin{equation}
 \beta(n)= -\frac{3 \zeta(3)}{16 \pi^2}  g_{v,s}m^2 \frac{d}{d n}\left \{ \frac{1}{n}\left [ \left (\frac{F_0^\rho}{1+F_0^\rho}\right )^2 + (g_{v,s}^2-1) \left (\frac{F_0^\sigma}{1+F_0^\sigma} \right )^2 \right ]\right \}  .
\label{eq:aa}
\end{equation}
In addition, in the case of weak short-ranged disorder, one needs to take into account in Eq. \eqref{eq:Sn:ff}
the renormalization of the effective mass \cite{Finkelstein1,Finkelstein2,Castellani1986,CA} due to combined effect of interaction and disorder: $m_* \to m_* z(T)$,
\begin{align}
z(T) & = 1 - \frac{1}{2\pi \sigma} \left[ \left ( \frac{F_0^\rho}{1+F_0^\rho}+ \frac{(g_{v,s}^2-1)F_0^\sigma}{1+F_0^\sigma} \right ) \ln \frac{E_F}{T} +  \frac{\ln(1+F_0^\rho) - a F_0^\rho}{1+F_0^\rho}+ (g_{v,s}^2-1) \frac{\ln(1+F_0^\sigma)-a F_0^\sigma}{1+F_0^\sigma} \right],
\label{eq:zz}
\end{align}
where $a = 1 - 3 [(2\gamma-3)\zeta(2)-\zeta^\prime(2)]/\pi^2 \approx 1.64$ (the Euler constant $\gamma\approx 0.577$)\cite{CA}. Here $\sigma = g_{v,s} E_F\tau$ denotes the dimensionless (in units of $e^2/h$) Drude conductivity. In the case of long range Coulomb interaction, one needs to take the limit $F_0^\rho\to\infty$ in Eqs. \eqref{eq:aa} and \eqref{eq:zz}.

~\\

{\large \bf Supplementary Note 9. Comparison of the thermodynamic and transport data.}
~\\

\begin{figure}
\begin{center}
\centerline{\psfig{figure=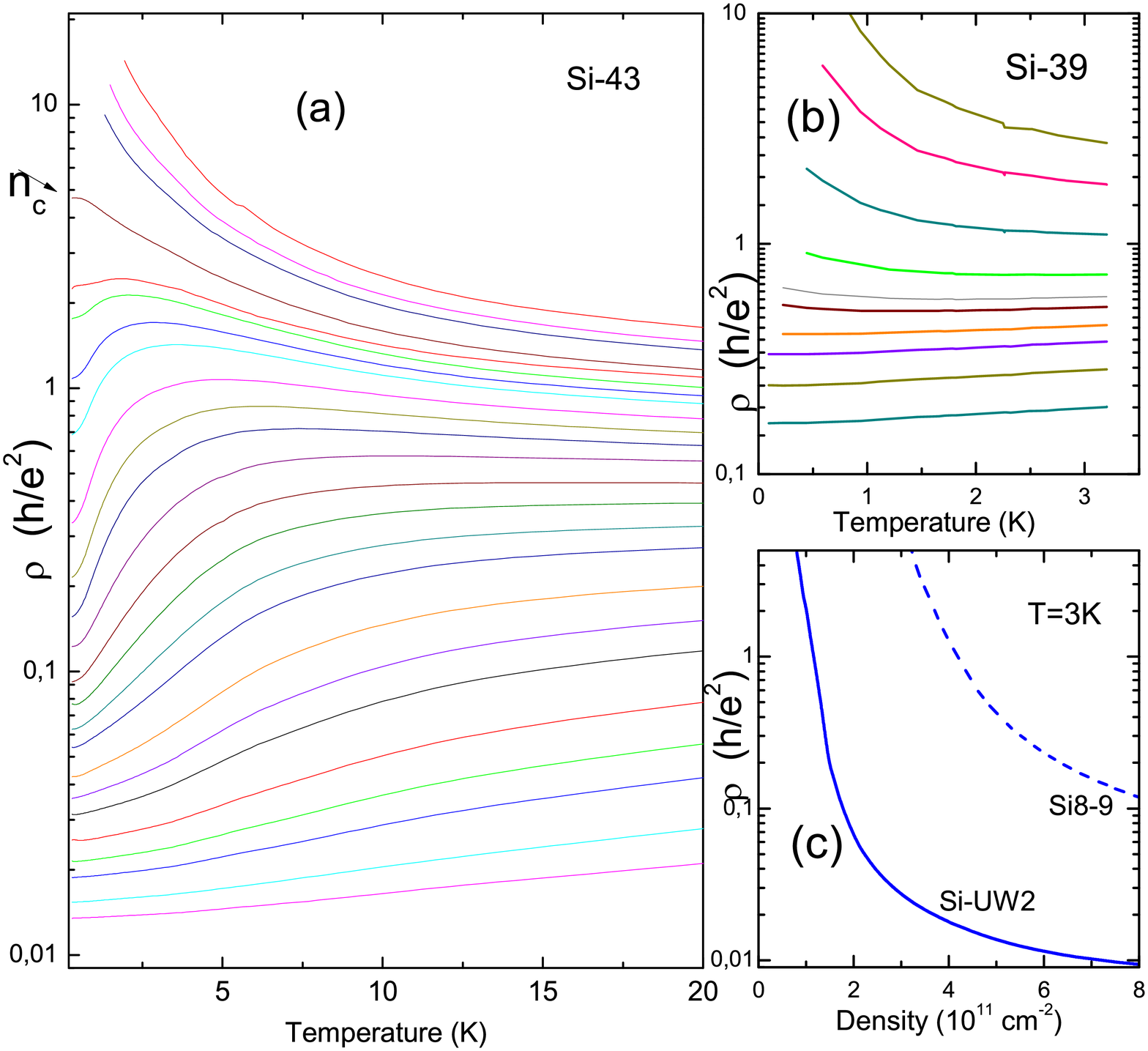, width=300pt}}
{{\bf Supplementary Figure 7. Resistivity data.} (a)Temperature dependence of resistivity for sample Si-43 (similar to Si-UW2 and Si-UW1) from Ref. \protect{\cite{AMP}} . Electron densities from bottom to top are(in units of $10^{11}$ cm$^{-2}$): 9.9, 8.12. 6.33, 5.43, 4.54,3.19,2.75,2.30, 2.07,1.85, 1.67, 1.49,1.34,1.25, 1.16, 1.075, 1.03, 0.985, 0.954, 0.895, 0.806, 0.770,  0.716. (b) Temperature dependence of resistivity for the disordered sample Si-39 (the same wafer as Si8-9) from Ref. \protect{\cite{AMP}} for various densities. Densities from bottom to top are (in units of $10^{11}$ cm$^{-2}$): 5.5,4.8,4.4,4.2,4.0,3.9,4.0,3.9,3.7,3.4,3.1,2.9. (c)Density dependence of the resistivity for samples Si-UW2 (solid line) and Si8-9 (dashed line) for a fixed temperature of 3K.}
\end{center}
\end{figure}

In zero magnetic field, the high-mobility Si-MOSFETs exhibit a metal-to-insulator transition (MIT). There are several theoretical  suggestions that this transition is a QPT, and several  experimental papers demonstrating scaling of various physical properties (transport, magnetization, thermo-EMF),   in support of this view. On the high density side of the transition there is a strong ''metallic`` ($d\rho/dT>0$) conductivity region. A reasonable question arises how the thermodynamic features reported here are related with the strong metallicity and MIT?

Using our high-temperature method we of course cannot address the QPT physics.


In Supplementary figures 7a and 7b we compare the temperature dependencies of the resistivity for the high mobility (clean)  and low mobility (disordered) samples. It is worth noting the low mobility MOSFETs which do not demonstrate the apparent MIT, also have metallic (though weak) temperature dependence of resistivity (see Refs.  \cite{AMP,PBr} for more detailed comparison of clean and disordered samples).
Comparison of the resistivity for two studied samples at a fixed temperature of 3K (see Supplementary Fig. 7c) clearly shows  that  the dissimilarity in their transport data (caused by different disorder) is irrelevant to the similarity in the entropy behavior  (see the main text).

~\\

{\large \bf Supplementary note 10.Testing the role of potential fluctuations.}
~\\

In the equilibrium, the  eleñtrochemical potential value is the same over the sample area. Inevitable (in high mobility samples) long-range disorder potential fluctuations force the local Fermi (kinetic) energy and the local density to fluctuate  and can even lead a system to microscopic phase separation. What we measure is  $\partial\langle\mu \rangle/\partial T$, averaged over the sample area. In this connection, a natural question arises: to what extent are the local potential fluctuation important?

Indeed, the upturn in the inverse compressibility \cite{allison, dultz} usually observed at low densities (close to the metal-insulator transition) is commonly accepted \cite{fogler, shi, ghoshal} as the firm evidence for the development of an inhomogeneous two-phase state.
Local compressibility measurements \cite{ilani1,ilani2} also directly demonstrate spatial inhomogeneity.

There are however several experimental arguments showing that in the explored range of parameters in our measurements the sample inhomogeneities are irrelevant:

\begin{itemize}

\item
Most of experimental indications of charge inhomogeneities were obtained at much lower temperatures. The local observations of the spatial non-uniformity in high-mobility GaAs samples  were done well below 1 K \cite{ilani1, ilani2}. Our magnetization measurements with high-mobility Si samples \cite{teneh,teneh2}  revealed the spin ``droplets'' with an enhanced spin (4 x single electron spin), persisting  in the 2D system on top of the Fermi liquid. These spin droplets however melt very rapidly with temperature (spin susceptibility behaves roughly as $\chi\propto T^{-2}$  \cite{teneh,teneh2}) and at $T\sim 3$ K the droplets have  negligible impact on thermodynamics. The absence of spin droplets is confirmed also by our entropy measurements at 3K in 5.5 Tesla parallel magnetic field  (see the corresponding paragraphs of the main text). There was a certain activity on transport detection of  droplet-related features in mesoscopic scale 2D systems \cite{ghosh,baenninger,goswami}. In all these papers signatures of droplets also arise only at low temperatures ($<1$ K).

\item
We note that the upturn of the inverse compressibility corresponds to reduction of the sample capacitance. In the above cited studies the onset of the inhomogeneous phase occurs close to or below $n_c$. The compressibility measurements in Refs.~\cite{allison, dultz} were done with GaAs- based systems.
With  our high mobility 2D electron system in  Si-UW2 sample we did not observe capacitance reduction over the whole explored range of densities, down to $n_0\approx 0.4\times 10^{11}$cm$^{-2}< n_c/2$, similar to that for n-GaAs in Ref.~\cite{allison}. For even lower densities the sample closes (i.e. fails to recharge), and its capacitance sharply drops down, impeding measurements for lower densities. The low mobility sample Si8-9 closes at higher densities $n\approx 1.5\times 10^{11}$\,cm$^{-2}$, however, it didn't manifest the MIT making senseless the question of the relation between the $n_0$ and $n_c$.

\item

As said above, an inhomogeneous sample state might set for $n<n_0$, where the capacitance sharply drops and the entropy measurements are inaccessible. Correspondingly, in the $dS/dn$ measurements (Fig.~2a of the main text), the data are missing in this range.
In order to fill this gap,  for simplicity, we have postulated $dS/dn= const$ for $n<n_0$.
 One might think that we loose an entropy related to this lowest density interval $0 - n_0$, when we integrate the measured signal from 0 to $n$. However, the integrated entropy (see insert to Fig.2d) for density $n=10.3\times 10^{11}$cm$^{-2}$ extrapolates to zero as temperature goes to zero, in accord with the 3rd Law of thermodynamics.  This result means that the inaccessible interval $0 - n_0$ does not contain notable amount of either excessive, or missing entropy. In other words, the entropy contribution from the localized (on inhomogeneities) phase is negligible and does not affect the measured $\langle S\rangle$ averaged over the sample area.

\item
The $n_0$ value also allows to make upper estimate of the screened potential fluctuation amplitude: $E_{fluct}<E_F(n_0)\approx 3K$ (otherwise, the hills of potential landscape exceed the Fermi level, the effective area of the sample and, hence, the capacitance start reducing). In a simple noninteracting picture the fluctuating dips of potential are responsible for band tail (see Supplementary Fig. 8a). We model the band tail within the simple Fermi-gas approach to see how it affects the signal.
Below we present results for two shapes of the band tail (a smooth one, Supplementary Fig. 8b, and the step-like one, Supplementary Fig. 8c).
\begin{figure}
\begin{center}
\centerline{\psfig{figure=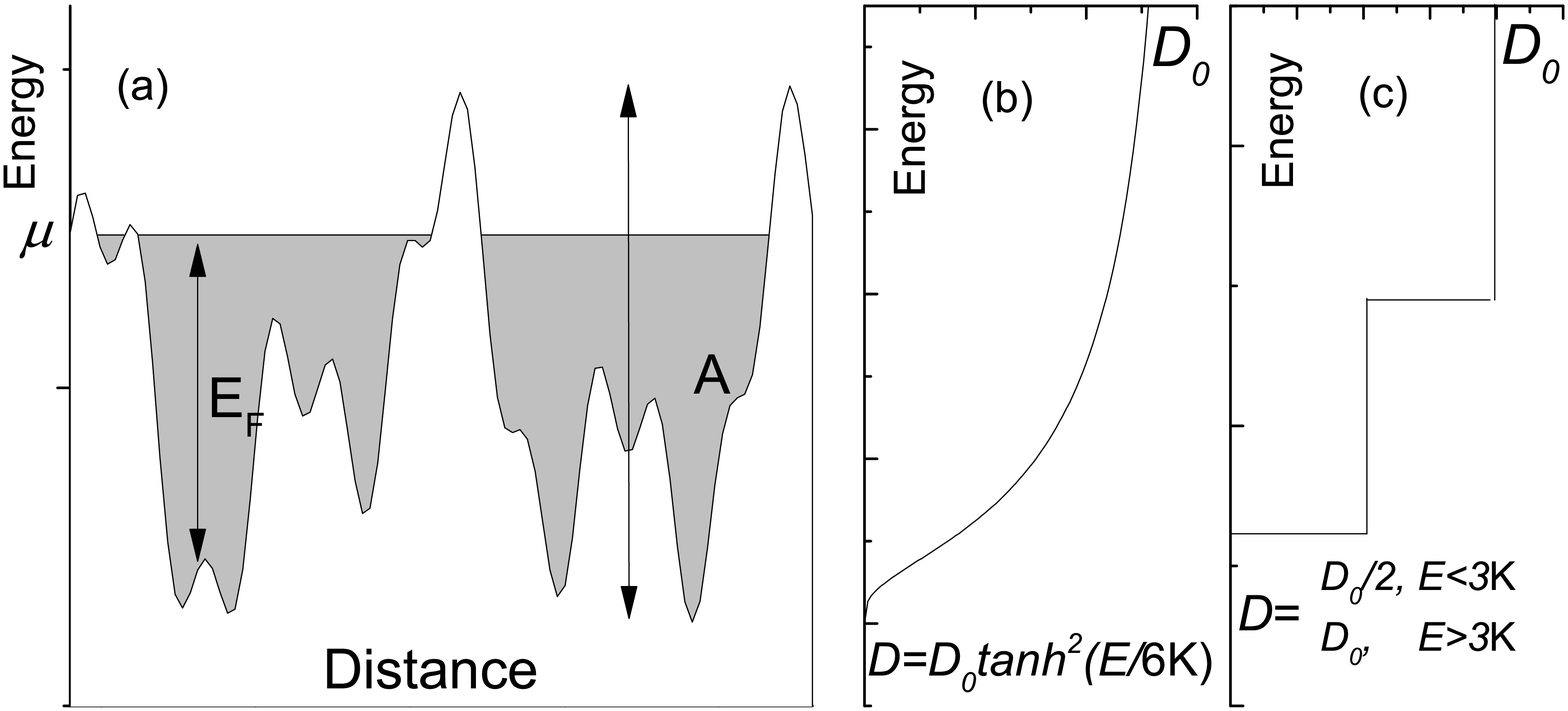, width=270pt}}
{{Supplementary Figure 8. Testing the role of semiclassical effect of disorder potential.}(a) Schematics of the spatial potential landscape. (b) The density of states with a smooth band tail (formula is given in the Figure). (c) The density of states with a step-like band tail.}
\end{center}
\end{figure}

\begin{figure}
\begin{center}
\centerline{\psfig{figure=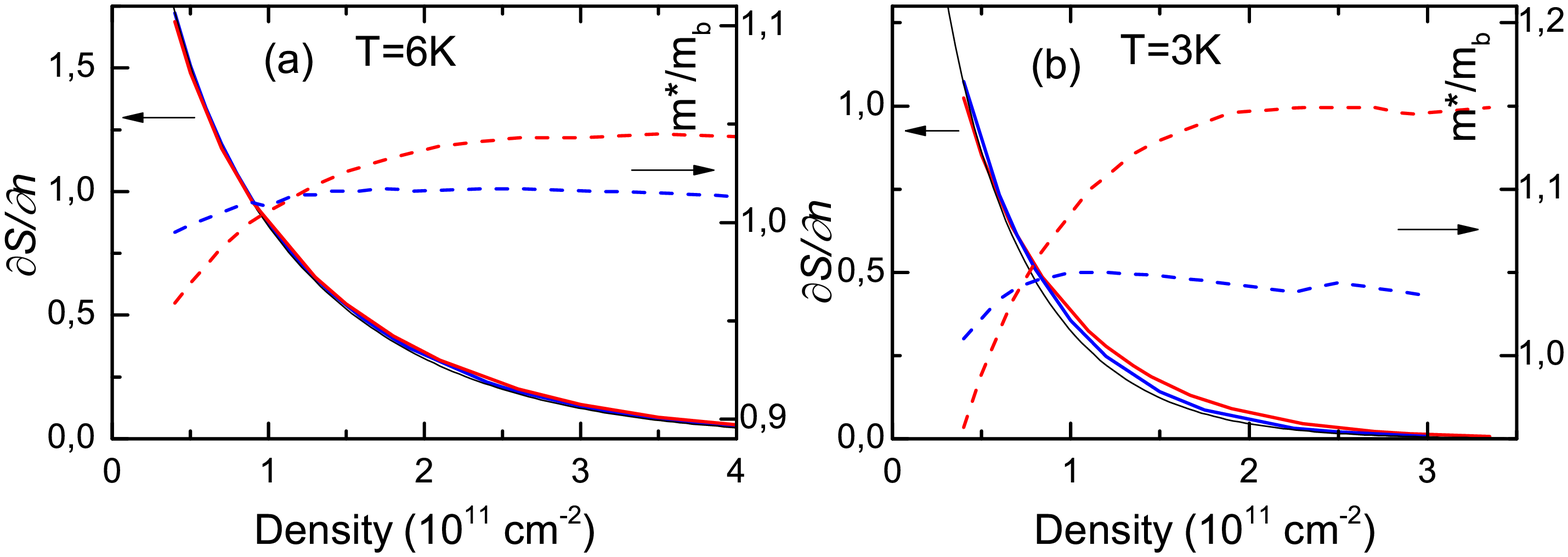, width=350pt}}
{{\bf Supplementary figure 9. Disorder impact on entropy per electron.} The $\partial S/\partial n (n)$ dependencies at different temperatures (a - 6K, b-3K) calculated for 2DEG in Si with tails in density of states: solid blue line $D(E)=D_0$, solid wine line $D(E)=D_0\tanh^2(E/5K)$, solid red line $D(E)$ is a two-step-like function(see Supplementary Figure 8). Dashed lines correspond to the right axes and show effective mass determined from the respective $\partial S/\partial n (n)$ data.}
\end{center}
\end{figure}

In order to find $\partial S/\partial n\equiv-\partial\mu/\partial T$ at given density and temperature and for given $D(E)$ dependence we solve numerically the equation with respect to $\mu$:
\begin{equation}
n=\int_0^\infty D(E)/(1+e^{(E-\mu)/T})dE.
\end{equation}
Further, we take a derivative $\partial\mu/\partial T$.
At $T=6$\,K (Supplementary Fig. 9a), the $dS/dn$ for 2D systems with and without the band  tail is almost indistinguishable.
The effective mass was determined from the data using Eq.~4 of the main text(i.e. in the same manner as it was done for the data processing). As a result, at $T=6$\,K we observe only a few percent difference in the effective mass determined for the model with and without the band tail. At $T=3$\,K (which is almost the lowest temperature in our experiments) the effective mass at higher densities exceeds the band value by up to $\sim 10\%$(Supplementary Fig. 9b) and tends to  drop  slightly for low densities.



\end{itemize}

The above consideration shows that within our constraints on nonuniformity of the system, and in the framework of the noninteracting model the thermodynamics of the non-degenerate Fermi gas at relatively high temperatures is not affected by potential fluctuations.
\begin{figure}
\begin{center}
\centerline{\psfig{figure=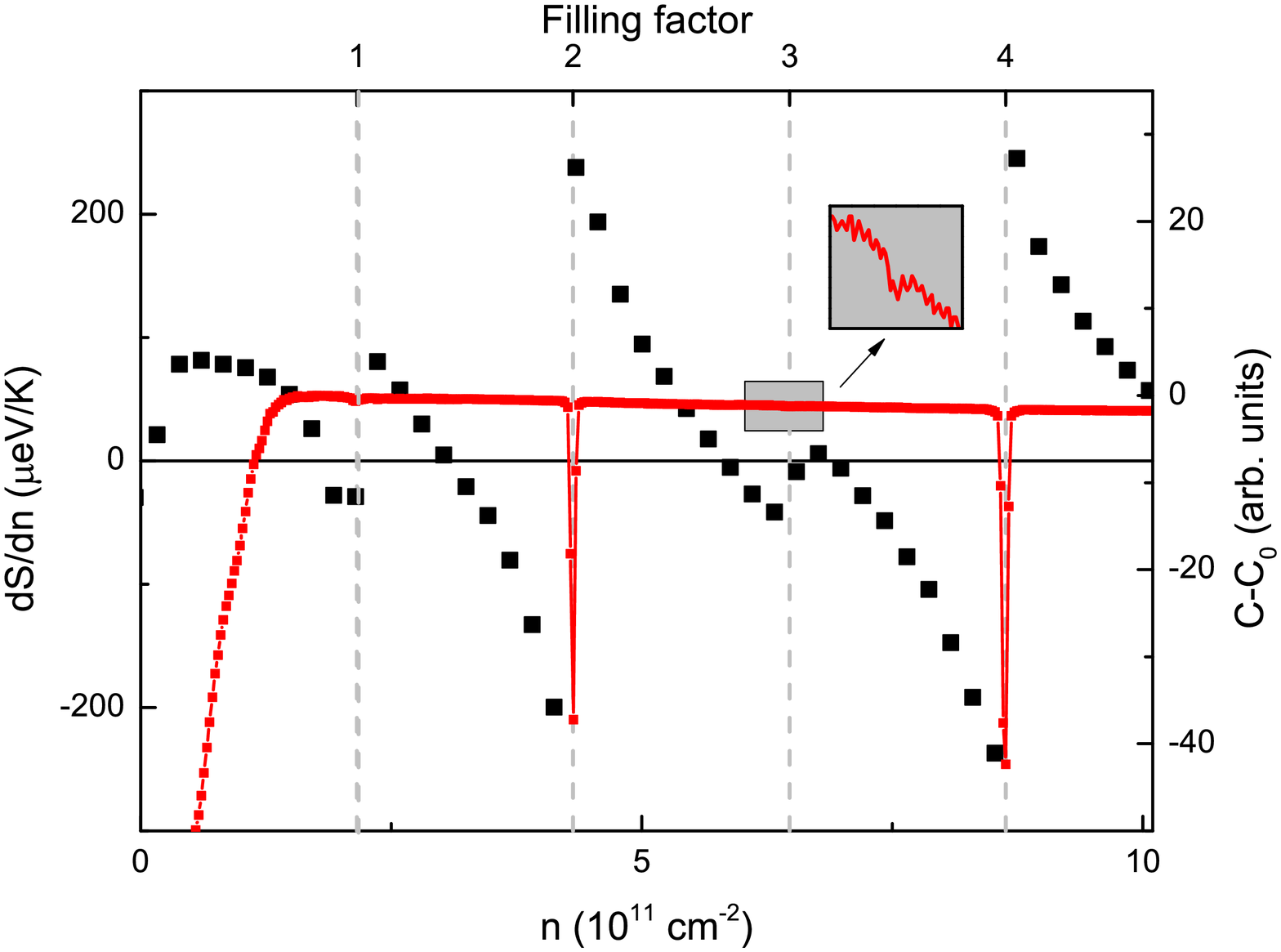, width=300pt}}
{{\bf Supplementary Figure 10. QHE features, used to determine the carrier density.} Capacitance measured at elevated frequency (red dots, right axis) and $\partial S/\partial n$ for Si-UW2 at 2.6K (black dots, left axis) in the QHE regime $B=9$ Tesla.}
\end{center}
\end{figure}

\begin{figure}
\begin{center}
\centerline{\psfig{figure=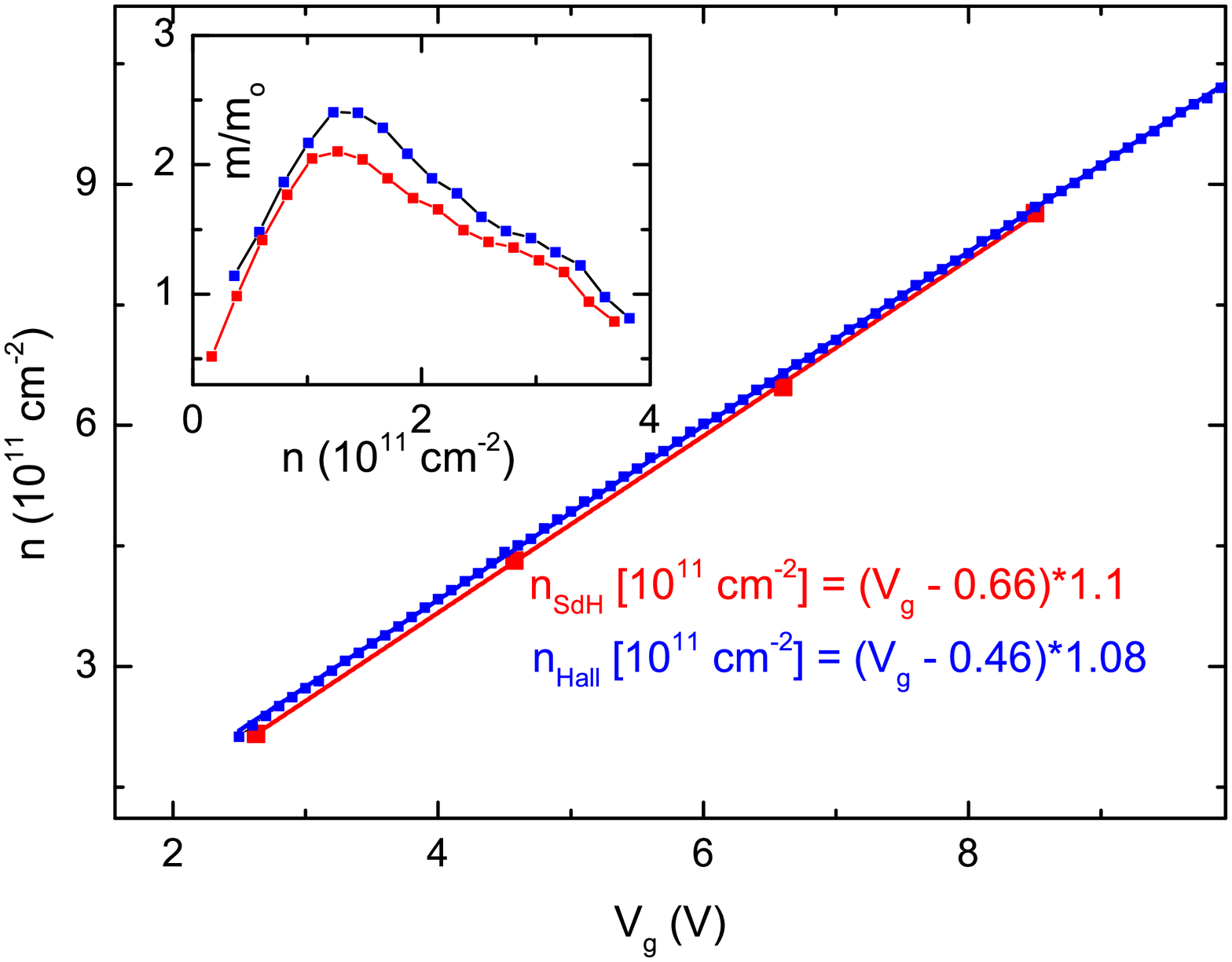, width=300pt}}
{{\bf Supplementary Figure 11. Two ways to determine electron density.}Electron density $n$ versus gate voltage $V_g$ for sample Si-UW2 at 2.6 K determined from Hall effect at 1 Tesla (blue dots) and from positions of QHE dips in capacitance at 9 Tesla using Eq.(\ref{density}) (red dots and line). Equations denote linear interpolations of the $n(V_g)$ dependencies. The inset shows the corresponding effective mass determined from Eq. 4 of the main text from $\partial S/\partial n$ data using both definitions of density for $T=4.8$K.}
\end{center}
\end{figure}

~\\

{\large \bf Supplementary note 11. Carrier density control.}
~\\

Being able to measure $\partial \mu/\partial T$, the capacitance, resistivity and the Hall resistivity we can estimate electron density and relate it to the features in thermodynamic  quantities. In the quantum Hall effect  (QHE) regime, capacitance sharply drops at integer filling factors:
\begin{equation}
n=\nu eB/2\pi\hbar.
\label{density}
\end{equation}
Due to these  sharp features, the electron density can be  quantified  in the QHE  regime of strong magnetic fields and high carrier densities, and, consequently, may be related to the gate voltage $V_g$,
$n=(C_0/e)(V_g-V_{\rm th})$, similar to that in the plane capacitor \cite{ando_review}.
As one can see from Supplementary Fig. 10, the capacitance drops exactly correspond to the  respective features in $\partial \mu/\partial T$. Information about electron density canalso be obtained from
the low-field Hall effect $R_{xy}=B/ne$ measured in both field directions with subsequent data antisymmetrization. For an ideal 2D system, both methods should give the same value for electron density. Practically, however, Hall density is usually a little bit higher \cite{JETPLett-Hall_2005}.

Both methods become inadequate in the insulating regime ($n<10^{11}$\,cm$^{-2}$) where quantization is suppressed, current flow becomes nonuniform and the ``Hall-insulator'' state often forms, in which the Hall voltage is irrelevant to the carrier density \cite{Hall_insulator}.
In this regime we can estimate electron density only by its extrapolation from higher densities. Supplementary Figure 11 shows comparison of the two methods for density determination. The insert to Supplementary Fig. 11 shows the effect of the chosen density definition  on the effective mass. It is seen that although definition of density for low $n$ affects effective mass considerably, it does not change qualitatively the $m(n)$ dependence and conclusions of our paper (see the main text).

\end{widetext}

\end{document}